\documentclass[journal,10pt,twoside,twocolumn]{IEEEtran}

\usepackage{cite}
\usepackage[T1]{fontenc}
\usepackage{graphicx}
\usepackage{amssymb}
\usepackage{amsmath}
\usepackage{amsthm}

\usepackage{paralist}

\usepackage{fancyref}
\usepackage{framed}
\usepackage{afterpage}
\usepackage{setspace}
\usepackage{microtype}
\usepackage{hyperref}
\usepackage{url}
\usepackage{booktabs}
\usepackage{algorithm}
\usepackage{algorithmic}
\usepackage{subfigure}
\usepackage{hyperref}
\usepackage{balance}
\usepackage[flushleft]{threeparttable}

\usepackage{amssymb}
\usepackage{amsfonts}
\usepackage{mathrsfs}
\usepackage{xspace}
\usepackage{bm}
\usepackage{upgreek}

\newcommand{\safemath}[2]{\newcommand{#1}{\ensuremath{#2}\xspace}}

\safemath{\bma}{\mathbf{a}}
\safemath{\bmb}{\mathbf{b}}
\safemath{\bmc}{\mathbf{c}}
\safemath{\bmd}{\mathbf{d}}
\safemath{\bme}{\mathbf{e}}
\safemath{\bmf}{\mathbf{f}}
\safemath{\bmg}{\mathbf{g}}
\safemath{\bmh}{\mathbf{h}}
\safemath{\bmi}{\mathbf{i}}
\safemath{\bmj}{\mathbf{j}}
\safemath{\bmk}{\mathbf{k}}
\safemath{\bml}{\mathbf{l}}
\safemath{\bmm}{\mathbf{m}}
\safemath{\bmn}{\mathbf{n}}
\safemath{\bmo}{\mathbf{o}}
\safemath{\bmp}{\mathbf{p}}
\safemath{\bmq}{\mathbf{q}}
\safemath{\bmr}{\mathbf{r}}
\safemath{\bms}{\mathbf{s}}
\safemath{\bmt}{\mathbf{t}}
\safemath{\bmu}{\mathbf{u}}
\safemath{\bmv}{\mathbf{v}}
\safemath{\bmw}{\mathbf{w}}
\safemath{\bmx}{\mathbf{x}}
\safemath{\bmy}{\mathbf{y}}
\safemath{\bmz}{\mathbf{z}}
\safemath{\bmzero}{\mathbf{0}}
\safemath{\bmone}{\mathbf{1}}

\bmdefine{\biad}{a}
\bmdefine{\bibd}{b}
\bmdefine{\bicd}{c}
\bmdefine{\bidd}{d}
\bmdefine{\bied}{e}
\bmdefine{\bifd}{f}
\bmdefine{\bigd}{g}
\bmdefine{\bihd}{h}
\bmdefine{\biid}{i}
\bmdefine{\bijd}{j}
\bmdefine{\bikd}{k}
\bmdefine{\bild}{l}
\bmdefine{\bimd}{m}
\bmdefine{\bind}{n}
\bmdefine{\biod}{o}
\bmdefine{\bipd}{p}
\bmdefine{\biqd}{q}
\bmdefine{\bird}{r}
\bmdefine{\bisd}{s}
\bmdefine{\bitd}{t}
\bmdefine{\biud}{u}
\bmdefine{\bivd}{v}
\bmdefine{\biwd}{w}
\bmdefine{\bixd}{x}
\bmdefine{\biyd}{y}
\bmdefine{\bizd}{z}

\bmdefine{\bixid}{\xi}
\bmdefine{\bilambdad}{\lambda}
\bmdefine{\bimud}{\mu}
\bmdefine{\bithetad}{\theta}
\bmdefine{\biphid}{\phi}
\bmdefine{\bideltad}{\delta}

\safemath{\bmia}{\biad}
\safemath{\bmib}{\bibd}
\safemath{\bmic}{\bicd}
\safemath{\bmid}{\bidd}
\safemath{\bmie}{\bied}
\safemath{\bmif}{\bifd}
\safemath{\bmig}{\bigd}
\safemath{\bmih}{\bihd}
\safemath{\bmii}{\biid}
\safemath{\bmij}{\bijd}
\safemath{\bmik}{\bikd}
\safemath{\bmil}{\bild}
\safemath{\bmim}{\bimd}
\safemath{\bmin}{\bind}
\safemath{\bmio}{\biod}
\safemath{\bmip}{\bipd}
\safemath{\bmiq}{\biqd}
\safemath{\bmir}{\bird}
\safemath{\bmis}{\bisd}
\safemath{\bmit}{\bitd}
\safemath{\bmiu}{\biud}
\safemath{\bmiv}{\bivd}
\safemath{\bmiw}{\biwd}
\safemath{\bmix}{\bixd}
\safemath{\bmiy}{\biyd}
\safemath{\bmiz}{\bizd}

\safemath{\bmxi}{\bixid}
\safemath{\bmlambda}{\bilambdad}
\safemath{\bmmu}{\bimud}
\safemath{\bmtheta}{\bithetad}
\safemath{\bmphi}{\biphid}
\safemath{\bmdelta}{\bideltad}

\safemath{\bA}{\mathbf{A}}
\safemath{\bB}{\mathbf{B}}
\safemath{\bC}{\mathbf{C}}
\safemath{\bD}{\mathbf{D}}
\safemath{\bE}{\mathbf{E}}
\safemath{\bF}{\mathbf{F}}
\safemath{\bG}{\mathbf{G}}
\safemath{\bH}{\mathbf{H}}
\safemath{\bI}{\mathbf{I}}
\safemath{\bJ}{\mathbf{J}}
\safemath{\bK}{\mathbf{K}}
\safemath{\bL}{\mathbf{L}}
\safemath{\bM}{\mathbf{M}}
\safemath{\bN}{\mathbf{N}}
\safemath{\bO}{\mathbf{O}}
\safemath{\bP}{\mathbf{P}}
\safemath{\bQ}{\mathbf{Q}}
\safemath{\bR}{\mathbf{R}}
\safemath{\bS}{\mathbf{S}}
\safemath{\bT}{\mathbf{T}}
\safemath{\bU}{\mathbf{U}}
\safemath{\bV}{\mathbf{V}}
\safemath{\bW}{\mathbf{W}}
\safemath{\bX}{\mathbf{X}}
\safemath{\bY}{\mathbf{Y}}
\safemath{\bZ}{\mathbf{Z}}

\safemath{\bZero}{\mathbf{0}}
\safemath{\bOne}{\mathbf{1}}
\safemath{\bDelta}{\mathbf{\Delta}}
\safemath{\bLambda}{\mathbf{\UpLambda}}
\safemath{\bPhi}{\mathbf{\Upphi}}
\safemath{\bSigma}{\mathbf{\Upsigma}}
\safemath{\bOmega}{\mathbf{\Upomega}}
\safemath{\bTheta}{\mathbf{\Uptheta}}

\bmdefine{\biAd}{A}
\bmdefine{\biBd}{B}
\bmdefine{\biCd}{C}
\bmdefine{\biDd}{D}
\bmdefine{\biEd}{E}
\bmdefine{\biFd}{F}
\bmdefine{\biGd}{G}
\bmdefine{\biHd}{H}
\bmdefine{\biId}{I}
\bmdefine{\biJd}{J}
\bmdefine{\biKd}{K}
\bmdefine{\biLd}{L}
\bmdefine{\biMd}{M}
\bmdefine{\biOd}{N}
\bmdefine{\biPd}{O}
\bmdefine{\biQd}{P}
\bmdefine{\biRd}{R}
\bmdefine{\biSd}{S}
\bmdefine{\biTd}{T}
\bmdefine{\biUd}{U}
\bmdefine{\biVd}{V}
\bmdefine{\biWd}{W}
\bmdefine{\biXd}{X}
\bmdefine{\biYd}{Y}
\bmdefine{\biZd}{Z}

\bmdefine{\biDelta}{\Delta}
\bmdefine{\biLambda}{\Lambda}
\bmdefine{\biPhi}{\Phi}
\bmdefine{\biSigma}{\Sigma}
\bmdefine{\biOmega}{\Omega}
\bmdefine{\biTheta}{\Theta}

\safemath{\bimA}{\biAd}
\safemath{\bimB}{\biBd}
\safemath{\bimC}{\biCd}
\safemath{\bimD}{\biDd}
\safemath{\bimE}{\biEd}
\safemath{\bimF}{\biFd}
\safemath{\bimG}{\biGd}
\safemath{\bimH}{\biHd}
\safemath{\bimI}{\biId}
\safemath{\bimJ}{\biJd}
\safemath{\bimK}{\biKd}
\safemath{\bimL}{\biLd}
\safemath{\bimM}{\biMd}
\safemath{\bimN}{\biNd}
\safemath{\bimO}{\biOd}
\safemath{\bimP}{\biPd}
\safemath{\bimQ}{\biQd}
\safemath{\bimR}{\biRd}
\safemath{\bimS}{\biSd}
\safemath{\bimT}{\biTd}
\safemath{\bimU}{\biUd}
\safemath{\bimV}{\biVd}
\safemath{\bimW}{\biWd}
\safemath{\bimX}{\biXd}
\safemath{\bimY}{\biYd}
\safemath{\bimZ}{\biZd}

\safemath{\bimDelta}{\biDelta}
\safemath{\bimLambda}{\biLambda}
\safemath{\bimPhi}{\biPhi}
\safemath{\bimSigma}{\biSigma}
\safemath{\bimOmega}{\biOmega}
\safemath{\bimTheta}{\biTheta}

\safemath{\setA}{\mathcal{A}}
\safemath{\setB}{\mathcal{B}}
\safemath{\setC}{\mathcal{C}}
\safemath{\setD}{\mathcal{D}}
\safemath{\setE}{\mathcal{E}}
\safemath{\setF}{\mathcal{F}}
\safemath{\setG}{\mathcal{G}}
\safemath{\setH}{\mathcal{H}}
\safemath{\setI}{\mathcal{I}}
\safemath{\setJ}{\mathcal{J}}
\safemath{\setK}{\mathcal{K}}
\safemath{\setL}{\mathcal{L}}
\safemath{\setM}{\mathcal{M}}
\safemath{\setN}{\mathcal{N}}
\safemath{\setO}{\mathcal{O}}
\safemath{\setP}{\mathcal{P}}
\safemath{\setQ}{\mathcal{Q}}
\safemath{\setR}{\mathcal{R}}
\safemath{\setS}{\mathcal{S}}
\safemath{\setT}{\mathcal{T}}
\safemath{\setU}{\mathcal{U}}
\safemath{\setV}{\mathcal{V}}
\safemath{\setW}{\mathcal{W}}
\safemath{\setX}{\mathcal{X}}
\safemath{\setY}{\mathcal{Y}}
\safemath{\setZ}{\mathcal{Z}}
\safemath{\emptySet}{\varnothing}

\safemath{\colA}{\mathscr{A}}
\safemath{\colB}{\mathscr{B}}
\safemath{\colC}{\mathscr{C}}
\safemath{\colD}{\mathscr{D}}
\safemath{\colE}{\mathscr{E}}
\safemath{\colF}{\mathscr{F}}
\safemath{\colG}{\mathscr{G}}
\safemath{\colH}{\mathscr{H}}
\safemath{\colI}{\mathscr{I}}
\safemath{\colJ}{\mathscr{J}}
\safemath{\colK}{\mathscr{K}}
\safemath{\colL}{\mathscr{L}}
\safemath{\colM}{\mathscr{M}}
\safemath{\colN}{\mathscr{N}}
\safemath{\colO}{\mathscr{O}}
\safemath{\colP}{\mathscr{P}}
\safemath{\colQ}{\mathscr{Q}}
\safemath{\colR}{\mathscr{R}}
\safemath{\colS}{\mathscr{S}}
\safemath{\colT}{\mathscr{T}}
\safemath{\colU}{\mathscr{U}}
\safemath{\colV}{\mathscr{V}}
\safemath{\colW}{\mathscr{W}}
\safemath{\colX}{\mathscr{X}}
\safemath{\colY}{\mathscr{Y}}
\safemath{\colZ}{\mathscr{Z}}

\safemath{\opA}{\mathbb{A}}
\safemath{\opB}{\mathbb{B}}
\safemath{\opC}{\mathbb{C}}
\safemath{\opD}{\mathbb{D}}
\safemath{\opE}{\mathbb{E}}
\safemath{\opF}{\mathbb{F}}
\safemath{\opG}{\mathbb{G}}
\safemath{\opH}{\mathbb{H}}
\safemath{\opI}{\mathbb{I}}
\safemath{\opJ}{\mathbb{J}}
\safemath{\opK}{\mathbb{K}}
\safemath{\opL}{\mathbb{L}}
\safemath{\opM}{\mathbb{M}}
\safemath{\opN}{\mathbb{N}}
\safemath{\opO}{\mathbb{O}}
\safemath{\opP}{\mathbb{P}}
\safemath{\opQ}{\mathbb{Q}}
\safemath{\opR}{\mathbb{R}}
\safemath{\opS}{\mathbb{S}}
\safemath{\opT}{\mathbb{T}}
\safemath{\opU}{\mathbb{U}}
\safemath{\opV}{\mathbb{V}}
\safemath{\opW}{\mathbb{W}}
\safemath{\opX}{\mathbb{X}}
\safemath{\opY}{\mathbb{Y}}
\safemath{\opZ}{\mathbb{Z}}
\safemath{\opZero}{\mathbb{O}}
\safemath{\identityop}{\opI}

\safemath{\veca}{\bma}
\safemath{\vecb}{\bmb}
\safemath{\vecc}{\bmc}
\safemath{\vecd}{\bmd}
\safemath{\vece}{\bme}
\safemath{\vecf}{\bmf}
\safemath{\vecg}{\bmg}
\safemath{\vech}{\bmh}
\safemath{\veci}{\bmi}
\safemath{\vecj}{\bmj}
\safemath{\veck}{\bmk}
\safemath{\vecl}{\bml}
\safemath{\vecm}{\bmm}
\safemath{\vecn}{\bmn}
\safemath{\veco}{\bmo}
\safemath{\vecp}{\bmp}
\safemath{\vecq}{\bmq}
\safemath{\vecr}{\bmr}
\safemath{\vecs}{\bms}
\safemath{\vect}{\bmt}
\safemath{\vecu}{\bmu}
\safemath{\vecv}{\bmv}
\safemath{\vecw}{\bmw}
\safemath{\vecx}{\bmx}
\safemath{\vecy}{\bmy}
\safemath{\vecz}{\bmz}

\safemath{\veczero}{\bmzero}
\safemath{\vecone}{\bmone}
\safemath{\vecxi}{\bmxi}
\safemath{\veclambda}{\bmlambda}
\safemath{\vecmu}{\bmmu}
\safemath{\vectheta}{\bmtheta}
\safemath{\vecphi}{\bmphi}
\safemath{\vecdelta}{\bmdelta}

\safemath{\matA}{\bA}
\safemath{\matB}{\bB}
\safemath{\matC}{\bC}
\safemath{\matD}{\bD}
\safemath{\matE}{\bE}
\safemath{\matF}{\bF}
\safemath{\matG}{\bG}
\safemath{\matH}{\bH}
\safemath{\matI}{\bI}
\safemath{\matJ}{\bJ}
\safemath{\matK}{\bK}
\safemath{\matL}{\bL}
\safemath{\matM}{\bM}
\safemath{\matN}{\bN}
\safemath{\matO}{\bO}
\safemath{\matP}{\bP}
\safemath{\matQ}{\bQ}
\safemath{\matR}{\bR}
\safemath{\matS}{\bS}
\safemath{\matT}{\bT}
\safemath{\matU}{\bU}
\safemath{\matV}{\bV}
\safemath{\matW}{\bW}
\safemath{\matX}{\bX}
\safemath{\matY}{\bY}
\safemath{\matZ}{\bZ}
\safemath{\matzero}{\bmzero}

\safemath{\matDelta}{\bDelta}
\safemath{\matLambda}{\bLambda}
\safemath{\matPhi}{\bPhi}
\safemath{\matSigma}{\bSigma}
\safemath{\matOmega}{\bOmega}
\safemath{\matTheta}{\bTheta}

\safemath{\matidentity}{\matI}
\safemath{\matone}{\matO}

\safemath{\rnda}{A}
\safemath{\rndb}{B}
\safemath{\rndc}{C}
\safemath{\rndd}{D}
\safemath{\rnde}{E}
\safemath{\rndf}{F}
\safemath{\rndg}{G}
\safemath{\rndh}{H}
\safemath{\rndi}{I}
\safemath{\rndj}{J}
\safemath{\rndk}{K}
\safemath{\rndl}{L}
\safemath{\rndm}{M}
\safemath{\rndn}{N}
\safemath{\rndo}{O}
\safemath{\rndp}{P}
\safemath{\rndq}{Q}
\safemath{\rndr}{R}
\safemath{\rnds}{S}
\safemath{\rndt}{T}
\safemath{\rndu}{U}
\safemath{\rndv}{V}
\safemath{\rndw}{W}
\safemath{\rndx}{X}
\safemath{\rndy}{Y}
\safemath{\rndz}{Z}

\safemath{\rveca}{\bimA}
\safemath{\rvecb}{\bimB}
\safemath{\rvecc}{\bimC}
\safemath{\rvecd}{\bimD}
\safemath{\rvece}{\bimE}
\safemath{\rvecf}{\bimF}
\safemath{\rvecg}{\bimG}
\safemath{\rvech}{\bimH}
\safemath{\rveci}{\bimI}
\safemath{\rvecj}{\bimJ}
\safemath{\rveck}{\bimK}
\safemath{\rvecl}{\bimL}
\safemath{\rvecm}{\bimM}
\safemath{\rvecn}{\bimN}
\safemath{\rveco}{\bomO}
\safemath{\rvecp}{\bimP}
\safemath{\rvecq}{\bimQ}
\safemath{\rvecr}{\bimR}
\safemath{\rvecs}{\bimS}
\safemath{\rvect}{\bimT}
\safemath{\rvecu}{\bimU}
\safemath{\rvecv}{\bimV}
\safemath{\rvecw}{\bimW}
\safemath{\rvecx}{\bimX}
\safemath{\rvecy}{\bimY}
\safemath{\rvecz}{\bimZ}

\safemath{\rvecxi}{\bmxi}
\safemath{\rveclambda}{\bmlambda}
\safemath{\rvecmu}{\bmmu}
\safemath{\rvectheta}{\bmtheta}
\safemath{\rvecphi}{\bmphi}

\safemath{\rmatA}{\bimA}
\safemath{\rmatB}{\bimB}
\safemath{\rmatC}{\bimC}
\safemath{\rmatD}{\bimD}
\safemath{\rmatE}{\bimE}
\safemath{\rmatF}{\bimF}
\safemath{\rmatG}{\bimG}
\safemath{\rmatH}{\bimH}
\safemath{\rmatI}{\bimI}
\safemath{\rmatJ}{\bimJ}
\safemath{\rmatK}{\bimK}
\safemath{\rmatL}{\bimL}
\safemath{\rmatM}{\bimM}
\safemath{\rmatN}{\bimN}
\safemath{\rmatO}{\bimO}
\safemath{\rmatP}{\bimP}
\safemath{\rmatQ}{\bimQ}
\safemath{\rmatR}{\bimR}
\safemath{\rmatS}{\bimS}
\safemath{\rmatT}{\bimT}
\safemath{\rmatU}{\bimU}
\safemath{\rmatV}{\bimV}
\safemath{\rmatW}{\bimW}
\safemath{\rmatX}{\bimX}
\safemath{\rmatY}{\bimY}
\safemath{\rmatZ}{\bimZ}

\safemath{\rmatDelta}{\bimDelta}
\safemath{\rmatLambda}{\bimLambda}
\safemath{\rmatPhi}{\bimPhi}
\safemath{\rmatSigma}{\bimSigma}
\safemath{\rmatOmega}{\bimOmega}
\safemath{\rmatTheta}{\bimTheta}

\usepackage{amssymb}
\usepackage{amsfonts}
\usepackage{mathrsfs}
\usepackage{xspace}
\usepackage{bm}
\usepackage{fancyref}
\usepackage{textcomp}

\usepackage{multirow}
\usepackage{stmaryrd}

\newenvironment{textbmatrix}{	\setlength{\arraycolsep}{2.5pt}%
								\big[\begin{matrix}}{\end{matrix}\big]%
								\raisebox{0.08ex}{\vphantom{M}}}

\def\be{\begin{equation}}
\def\ee{\end{equation}}
\def\een{\nonumber \end{equation}}
\def\mat{\begin{bmatrix}}
\def\emat{\end{bmatrix}}
\def\btm{\begin{textbmatrix}}
\def\etm{\end{textbmatrix}}

\def\ba#1\ea{\begin{align}#1\end{align}}
\def\bas#1\eas{\begin{align*}#1\end{align*}}
\def\bs#1\es{\begin{split}#1\end{split}} 
\def\bg#1\eg{\begin{gather}#1\end{gather}}
\def\bml#1\eml{\begin{multline}#1\end{multline}}
\def\bi#1\ei{\begin{itemize}#1\end{itemize}}

\newcommand{\lefto}{\mathopen{}\left}

\DeclareMathOperator{\sign}{sgn}			
\DeclareMathOperator*{\argmin}{arg\;min}		
\DeclareMathOperator{\kron}{\otimes}			
\DeclareMathOperator{\Exop}{\opE}			

\newcommand{\abs}[1]{\lefto\lvert#1\right\rvert}		

\newcommand{\vecnorm}[1]{\lefto\lVert#1\right\rVert}		

\safemath{\dirac}{\delta}					
\safemath{\krond}{\dirac}					

\safemath{\upto}{\uparrow}
\safemath{\downto}{\downarrow}
\safemath{\iu}{j}							
\safemath{\ev}{\lambda}						
\safemath{\hilseqspace}{l^{2}}				
\newcommand{\banachfunspace}[1]{\setL^{#1}}	
\safemath{\hilfunspace}{\banachfunspace{2}}	

\safemath{\SNR}{\textsf{SNR}} 				
\safemath{\PAR}{\textsf{PAR}} 				
\safemath{\No}{N_0}							
\safemath{\Es}{E_s}							
\safemath{\Eb}{E_b}							
\safemath{\EbNo}{\frac{\Eb}{\No}}
\safemath{\EsNo}{\frac{\Es}{\No}}

\DeclareMathOperator{\CHop}{\ensuremath{\opH}} 
\safemath{\tvir}{\rndh_{\CHop}}				
\safemath{\tvtf}{\rndl_{\CHop}}				
\safemath{\spf}{\rnds_{\CHop}}				
\safemath{\bff}{H_{\CHop}}					

\safemath{\ircf}{r_{h}}						
\safemath{\tftvcf}{r_{s}}					
\safemath{\tfcf}{r_{l}}						
\safemath{\bfcf}{r_{H}}						

\safemath{\tcorr}{c_h}						
\safemath{\scf}{c_{s}}						
\safemath{\tfcorr}{c_{l}}					
\safemath{\fcorr}{c_{H}}						

\safemath{\mi}{I}							
\safemath{\capacity}{C}						

\safemath{\normal}{\mathcal{N}}			
\safemath{\jpg}{\mathcal{CN}}			
\safemath{\mchain}{\leftrightarrow}		

\safemath{\dB}{\,\mathrm{dB}}
\safemath{\dBm}{\,\mathrm{dBm}}
\safemath{\Hz}{\,\mathrm{Hz}}
\safemath{\kHz}{\,\mathrm{kHz}}
\safemath{\MHz}{\,\mathrm{MHz}}
\safemath{\GHz}{\,\mathrm{GHz}}
\safemath{\s}{\,\mathrm{s}}
\safemath{\ms}{\,\mathrm{ms}}
\safemath{\mus}{\,\mathrm{\text{\textmu}s}}
\safemath{\ns}{\,\mathrm{ns}}
\safemath{\ps}{\,\mathrm{ps}}
\safemath{\meter}{\,\mathrm{m}}
\safemath{\mm}{\,\mathrm{mm}}
\safemath{\cm}{\,\mathrm{cm}}
\safemath{\m}{\,\mathrm{m}}
\safemath{\W}{\,\mathrm{W}}
\safemath{\mW}{\, \mathrm{mW}}
\safemath{\J}{\,\mathrm{J}}
\safemath{\K}{\,\mathrm{K}}
\safemath{\bit}{\,\mathrm{bit}}
\safemath{\nat}{\,\mathrm{nat}}

\safemath{\define}{\triangleq}			

\safemath{\equivalent}{\sim}
\safemath{\distas}{\sim}					
\safemath{\sdiff}{\Delta}				

\safemath{\reals}{\mathbb{R}}
\safemath{\positivereals}{\reals_{+}}
\safemath{\integers}{\mathbb{Z}}
\safemath{\posint}{\integers_{+}}
\safemath{\naturals}{\mathbb{N}}
\safemath{\posnaturals}{\naturals_{+}}
\safemath{\complexset}{\mathbb{C}}
\safemath{\rationals}{\mathbb{Q}}

\newcommand*{\fancyrefapplabelprefix}{app}		
\newcommand*{\fancyrefthmlabelprefix}{thm}		
\newcommand*{\fancyreflemlabelprefix}{lem}		
\newcommand*{\fancyrefcorlabelprefix}{cor}		
\newcommand*{\fancyrefdeflabelprefix}{def}		
\newcommand*{\fancyrefproplabelprefix}{prop}	
\newcommand*{\fancyrefobslabelprefix}{obs}		
\newcommand*{\fancyrefalglabelprefix}{alg}		
\newcommand*{\fancyrefasmlabelprefix}{asm}	    
\newcommand*{\fancyreftbllabelprefix}{tbl}	    

\frefformat{vario}{\fancyrefseclabelprefix}{Section~#1}
\frefformat{vario}{\fancyrefthmlabelprefix}{Theorem~#1}
\frefformat{vario}{\fancyreflemlabelprefix}{Lemma~#1}
\frefformat{vario}{\fancyrefcorlabelprefix}{Corollary~#1}
\frefformat{vario}{\fancyrefdeflabelprefix}{Definition~#1}
\frefformat{vario}{\fancyrefobslabelprefix}{Observation~#1}
\frefformat{vario}{\fancyrefasmlabelprefix}{Assumption~#1}
\frefformat{vario}{\fancyreffiglabelprefix}{Figure~#1}
\frefformat{vario}{\fancyrefapplabelprefix}{Appendix~#1} 
\frefformat{vario}{\fancyrefproplabelprefix}{Proposition~#1}
\frefformat{vario}{\fancyrefalglabelprefix}{Algorithm~#1}
\frefformat{vario}{\fancyrefeqlabelprefix}{(#1)}
\frefformat{vario}{\fancyreftbllabelprefix}{Table~#1}

\newtheorem{thm}{Theorem}

\newtheorem{lem}[thm]{Lemma} 

\newtheorem{rem}{Remark}

\safemath{\dictab}{[\,\dicta\,\,\dictb\,]}

\safemath{\ysig}{\bmy}
\safemath{\ysighat}{\hat{\ysig}}
\safemath{\ysigdim}{M}
\safemath{\xsig}{\bmx}
\safemath{\xsigdim}{N}
\safemath{\nx}{n_x}
\safemath{\zsig}{\bmz}
\safemath{\zsigdim}{\ysigdim}
\safemath{\rsig}{\bmr}
\safemath{\Adict}{\bA}
\safemath{\Adicttilde}{\widetilde{\Adict}}
\safemath{\Adictdim}{\outputdim\times\xsigdim}
\safemath{\avec}{\bma}
\safemath{\avectilde}{\tilde{\avec}}
\safemath{\Bdict}{\bB}
\safemath{\Bdicttilde}{\widetilde{\Bdict}}
\safemath{\Cdict}{\bC}
\safemath{\cvec}{\bmc}
\safemath{\Ddict}{\bD}
\safemath{\Ddictdim}{\ysigdim\times\xsigdim}
\safemath{\dvec}{\bmd}
\safemath{\Ddicttilde}{\widetilde{\bD}}
\safemath{\Bonb}{\bB}
\safemath{\bvec}{\bmb}
\safemath{\Bonbdim}{\ysigdim\times\ysigdim}
\safemath{\noise}{\bmn}
\safemath{\noisedim}{\ysigim}
\safemath{\err}{\bme}
\safemath{\errdim}{\ysigdim}
\safemath{\errset}{\setE}
\safemath{\nerr}{n_e}
\safemath{\delop}{\bP_\errset}
\safemath{\delopc}{\bP_{{\errset}^c}}

\safemath{\cplxi}{\imath}
\safemath{\cplxj}{\jmath}

\safemath{\dict}{\matD}
\safemath{\inputdim}{N}		
\safemath{\outputdim}{M}		
\safemath{\sparsity}{S}	
\safemath{\inputdimA}{{N_a}}	
\safemath{\inputdimB}{{N_b}}	
\safemath{\elemA}{{n_a}}	
\safemath{\elemB}{{n_b}}	
\safemath{\resA}{\matR_a}	
\safemath{\resB}{\matR_b}	
\safemath{\subD}{\matS} 
\safemath{\subA}{\matS_a} 
\safemath{\subB}{\matS_b} 
\safemath{\dicta}{\matA} 	
\safemath{\dictb}{\matB} 	
\safemath{\hollowS}{H}
\safemath{\hollowA}{H_a}
\safemath{\hollowB}{H_b}
\safemath{\cross}{Z}
\safemath{\coh}{\mu_d}			
\safemath{\coha}{\mu_a}			
\safemath{\cohb}{\mu_b}			
\safemath{\mubs}{\nu}	
\safemath{\cohm}{\mu_m} 
\safemath{\dictset}{\setD}	
\safemath{\dictsetp}{\dictset(\coh,\coha,\cohb)}	
\safemath{\dictsetgen}{\dictset_\text{gen}}
\safemath{\dictsetgenp}{\dictsetgen(\coh)}
\safemath{\dictsetonb}{\dictset_\text{onb}}
\safemath{\dictsetonbp}{\dictsetonb(\coh)}

\safemath{\leftside}{U}
\safemath{\rightsideA}{R_a}
\safemath{\rightsideB}{R_b}

\safemath{\indexS}{\setI_S} 

\safemath{\na}{n_a}			
\safemath{\nb}{n_b}			
\safemath{\coeffa}{p_i}	
\safemath{\coeffb}{q_j}	
\safemath{\seta}{\setP}		
\safemath{\setb}{\setQ}     
\safemath{\setw}{\setW}	
\safemath{\setz}{\setZ}	
\safemath{\cola}{\veca}		
\safemath{\colb}{\vecb}		
\safemath{\cold}{\vecd}		
\safemath{\inputvec}{\vecx} 	
\safemath{\error}{\vece}	
\safemath{\noiseout}{\vecz} 	
\safemath{\inputvecel}{x}
\safemath{\inputveca}{\vecx_a}
\safemath{\inputvecb}{\vecx_b}
\safemath{\outputvec}{\vecy}	
\safemath{\lambdamin}{\lambda_{\mathrm{min}}}

\safemath{\elltwo}{\ell_2}
\safemath{\ellone}{\ell_1}
\safemath{\ellzero}{\ell_0}
\safemath{\ellinf}{\ell_\infty}
\safemath{\ellinftilde}{\ell_{\widetilde\infty}}
\safemath{\licard}{Z(\coh,\coha,\cohb)}
\safemath{\xsol}{\hat{x}}
\safemath{\xbord}{x_b}		
\safemath{\xstat}{x_s}		
\safemath{\xstatLone}{\tilde{x}_s}
\safemath{\order}{\mathcal{O}} 
\safemath{\scales}{\Theta} 
\safemath{\ones}{\mathbf{1}} 
\safemath{\zeroes}{\mathbf{0}} 
\safemath{\thlone}{\kappa(\coh,\cohb)} 
\safemath{\constoneA}{\delta} 
\safemath{\constoneB}{\epsilon} 
\safemath{\nlarge}{L}				   
\safemath{\sumlarge}{S_\nlarge}
\safemath{\maxlarger}{P_\nlarge}	   
\safemath{\Pzero}{\textrm{P0}}	
\safemath{\Pone}{\textrm{P1}}
\safemath{\vecfir}{\vecw}			 
\safemath{\vecsec}{\vecz}
\safemath{\elvecfir}{w}              
\safemath{\elvecsec}{z}				 
\safemath{\nlargefir}{n}
\safemath{\normout}{\gamma}
\safemath{\auxfun}{h}
\safemath{\supp}{\textrm{supp}}

\safemath{\indexa}{\ell}
\safemath{\indexb}{r}
\safemath{\indexc}{i}
\safemath{\indexd}{j}

\safemath{\project}{P}

\IEEEoverridecommandlockouts 

\allowdisplaybreaks

\usepackage{color}
\newcommand{\revision}[1]{#1}

\begin{document}
\title{Decentralized Baseband Processing \\ for Massive MU-MIMO Systems}

\author{Kaipeng Li,  Rishi Sharan,   Yujun Chen, Tom Goldstein, Joseph R. Cavallaro, and  Christoph Studer

\thanks{K.~Li, Y.~Chen, and J.~R.~Cavallaro are with the Department of Electrical and Computer Engineering, Rice University, Houston 77251, TX (e-mail: kl33@rice.edu; yujun.chen@rice.edu; cavallar@rice.edu).}
\thanks{R.~Sharan was with the School of Electrical and Computer Engineering, Cornell University, Ithaca 14853, NY, and is now at The MITRE Corporation, McLean 22102, VA (e-mail: rrs72@cornell.edu).} 
\thanks{T.~Goldstein is with the Department of Computer Science, University of Maryland, College Park 20740, MD (e-mail: tomg@cs.umd.edu).}
\thanks{C.~Studer is  with the School of Electrical and Computer Engineering, Cornell University, Ithaca 14853, NY (e-mail: studer@cornell.edu; web: \url{http://vip.ece.cornell.edu}; corresponding author).}
\thanks{Parts of this paper have been presented at the 2016 GlobalSIP Conference~\cite{admm_gpu} and the  Asilomar Conference on Signals, Systems, and Computers \cite{admm_xeon_phi}. The present paper contains a new ADMM-based data detection algorithm and a generalized ADMM-based beamforming algorithm, as well as corresponding reference implementations on a GPU cluster for the uplink and downlink.}
\thanks{\revision{A MATLAB simulator for decentralized baseband processing as proposed in this paper is available on GitHub: \url{https://github.com/VIP-Group/DBP}}}
}

\maketitle


\begin{abstract}
Achieving high spectral efficiency in realistic massive multi-user (MU) multiple-input multiple-output (MIMO) wireless systems requires computationally-complex algorithms for data detection in the uplink (users transmit to {base-station}) and beamforming in the downlink ({base-station} transmits to users).
Most existing algorithms are designed to be executed on centralized computing hardware at the {base-station} (BS), {which results in prohibitive  complexity} for systems with hundreds or thousands of antennas and generates raw baseband data rates that exceed the limits of current interconnect technology and chip I/O interfaces. 
This paper proposes  a novel decentralized baseband processing architecture that alleviates these bottlenecks by partitioning the BS antenna array into clusters, each associated with independent radio-frequency chains, analog and digital modulation circuitry, and computing hardware.
For this architecture, we develop novel decentralized data detection and beamforming algorithms that only access local channel-state information and require low communication bandwidth among the clusters. 
We study the associated trade-offs between error-rate performance, computational complexity, and interconnect bandwidth, and we demonstrate the scalability of our solutions for massive MU-MIMO systems with  thousands of BS antennas using reference implementations on a graphic processing unit~(GPU) cluster. 
\end{abstract}



\section{Introduction}

\IEEEPARstart{M}{assive} multi-user (MU) multiple-input multiple-output (MIMO) is among the most promising technologies  for realizing high spectral efficiency and improved link reliability in fifth-generation (5G) wireless systems~\cite{mimo_overview, mimo_next_gen}. The main idea behind massive MU-MIMO is to equip the base-station (BS) with hundreds or thousands of antenna elements, which increases the spatial resolution and provides an energy-efficient way to serve a large number of users in the same time-frequency resource. 
Despite all the advantages of this emerging technology, the presence of a large number of BS antenna elements results in a variety of  implementation challenges. 
One of the most critical challenges is the excessively high amount of raw baseband data that must be transferred from the baseband processing unit to the radio-frequency (RF) antenna units at the BS (or in the opposite direction). 
Consider, for example, a $128$ BS-antenna massive MU-MIMO system with $40$\,MHz bandwidth and $10$-bit analog-to-digital converters (ADCs). For such a system, the raw baseband data rates from and to the RF units easily exceed $200$\,Gbit/s.
Such high data rates not only pose severe implementation challenges for the computing hardware to carry out the necessary baseband processing tasks, but the resulting raw baseband data stream may also exceed the bandwidth of existing high-speed interconnects, such as the common public radio interface (CPRI)~\cite{cpri}.

\setlength{\textfloatsep}{5pt}
\begin{figure*}[t]
\centering
\includegraphics[width=0.95\textwidth]{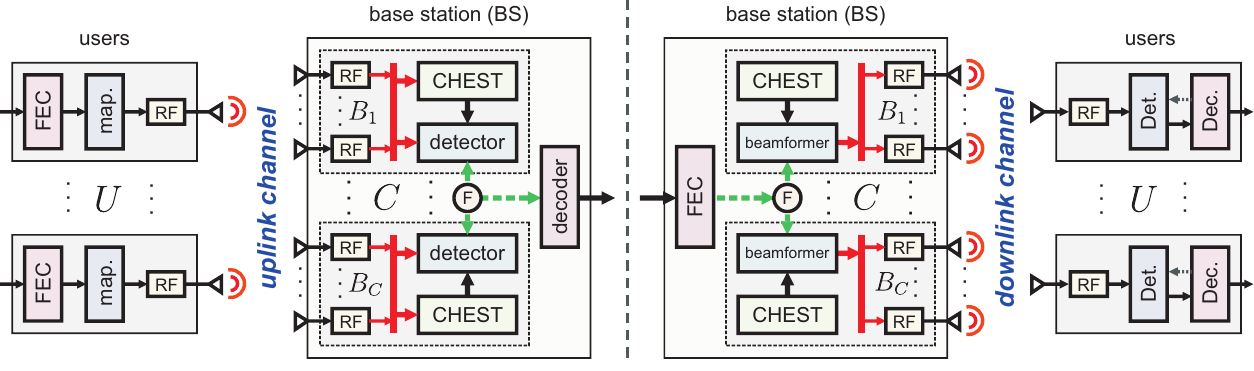}
\caption{Overview of the proposed decentralized baseband processing (DBP) architecture. Left: Massive MU-MIMO uplink: $U$ single-antenna users communicate to the base-station (BS). The $B$ BS antenna elements are divided into $C$ clusters, which independently perform channel estimation (CHEST) and decentralized data detection. Right: Massive MU-MIMO downlink: The BS performs decentralized beamforming; each of the $C$ clusters only uses local channel state information. In both scenarios, only a minimum amount of consensus information is exchanged among the clusters (indicated by the dashed green lines).}
\label{fig:system_overview}
\vspace{-0.4cm}
\end{figure*}

\subsection{Challenges of Centralized Baseband Processing}
Recent testbed implementations for massive MU-MIMO, such as the Argos testbed~\cite{argos,argosv2}, the LuMaMi testbed~\cite{lund}, and the BigStation~\cite{bigstation}, reveal that the centralized baseband processing required for data detection in the uplink (users communicate to BS) and downlink (BS  communicates to users using beamforming) is extremely challenging with current interconnect technology.
In fact, all of the  proposed data detection or beamforming algorithms that realize the full benefits of massive MU-MIMO systems with realistic (finite) antenna configurations, such as zero-forcing (ZF) or minimum mean-square error (MMSE) equalization or beamforming~\cite{howmany}, rely on \emph{centralized baseband processing}. {This approach requires that full channel state information (CSI) and all receive/transmit data streams and all subcarriers (for wideband systems) are available at a centralized node, which processes and generates the baseband signals that are received from and transmitted to the radio-frequency (RF) chains.}
To avoid such a traditional, centralized baseband processing approach, existing testbeds, such as the Argos testbed~\cite{argos}, rely on maximum-ratio combining (MRC), which enables fully decentralized channel estimation, data detection, and beamforming directly at the antenna elements.
Unfortunately, MRC significantly reduces the spectral efficiency for realistic antenna configurations  compared to that of ZF or MMSE-based methods~\cite{howmany}, which prevents the use of high-rate modulation and coding schemes that fully exploit the advantages of massive MU-MIMO.

\subsection{Decentralized Baseband Processing (DBP)}
\label{sec:DBPoverview}

In this paper, we propose a \emph{decentralized baseband processing (DBP)} architecture as illustrated in \fref{fig:system_overview}, which alleviates the bottlenecks of massive MU-MIMO caused by extremely high raw baseband data rates and implementation complexity of centralized processing.\footnote{{For the sake of simplicity, the BS illustration in \fref{fig:system_overview} only includes the components that are relevant to the results in this paper.}}
We partition the $B$ BS antennas into~$C$ independent clusters, each having $B_c$ antennas for the $c$th cluster so that $B=\sum_{c=1}^C B_c$. 
For simplicity, we will assume clusters of equal size and set $S=B_c$ which implies $SC=B$.
{Each cluster is associated with local computing hardware, a so-called \emph{processing element (PE)}, that carries out the necessary baseband processing tasks in a decentralized and parallel fashion.} A central fusion node (``\textsf{F}'' in  \fref{fig:system_overview}) processes a small amount of consensus information  that is exchanged among the clusters and required by our decentralized baseband algorithms (the dashed green lines  in  \fref{fig:system_overview}). 

Throughput the paper, we focus on time-division duplexing~(TDD), i.e., we alternate between uplink and downlink communication  within the same frequency band. 
In the uplink phase, $U$ users communicate with the BS. First, CSI is acquired via pilots at the BS and stored locally at each cluster. Then, data is transmitted by the users and decoded at the BS.  
In the downlink phase, the BS transmits data to the~$U$ users. By exploiting channel reciprocity, the BS performs decentralized beamforming (or precoding) to mitigate MU interference (MUI) and to focus transmit power towards the users. As for the uplink, the $C$ decentralized beamforming units only access local CSI. 

The key features of the proposed DBP architecture can be summarized as follows:
(i)  DBP reduces the raw baseband data rates between each cluster and the associated RF chains. In addition, the I/O bandwidth of each PE can be reduced significantly as only raw baseband data from a (potentially small) subset of antennas must be transferred on and off chip.
(ii) {DBP lowers the computational complexity per PE by distributing and parallelizing the key signal-processing tasks. In addition to decentralizing channel estimation (CHEST), data detection, and beamforming (or precoding), DBP enables frequency-domain processing} (e.g., fast Fourier transforms for orthogonal frequency-division multiplexing) as well as impairment compensation (e.g., for carrier frequency and sampling rate offsets, phase noise, or I/Q imbalance) locally at each cluster.
(iii) DPB enables modular and scalable BS designs; adding or removing antenna elements simply amounts to adding or removing computing clusters and the associated RF elements, respectively. 
(iv) DPB allows one to distribute the antenna array and the associated computing hardware over multiple buildings---an idea that was put forward recently in the massive MU-MIMO context~\cite{niwhite}. 

\subsection{Relevant Prior Art}
{The literature describes mainly three methods that are related to DBP: coordinated multipoint (CoMP), cloud radio access networks (C-RAN), and testbeds that perform distributed baseband processing across frequencies (or subcarriers).} The following paragraphs discuss these results.

\subsubsection{Coordinated multipoint (CoMP)} 
Coordinated multipoint (CoMP)  is a distributed communication technology to eliminate inter-cell interference,  improve the data rate, and increase the spectrum efficiency for cell-edge users~\cite{compconcept}. 
CoMP distributes multiple BSs across cells, which cooperate via backhaul interconnect to perform distributed uplink reception and downlink transmission.
CoMP has been studied for cooperative transmission  and reception~\cite{complte1,complte2}  in 3GPP LTE-A, and is widely believed to play an important role in 5G networks~\cite{comp5g} along with other technologies, such as massive MU-MIMO~\cite{compmimo}.
Several algorithms for distributed beamforming with CoMP have been proposed in~\cite{dpc_precoding,gs_precoding,admm_mcbf}. 
The paper \cite{dpc_precoding} proposes a distributed precoding algorithm for multi-cell MIMO downlink systems using a dirty-paper coding.
The papers \cite{gs_precoding,admm_mcbf} propose distributed beamforming algorithms based  on Gauss-Seidel and  alternating direction method of multipliers (ADMM).
These methods assume that the BSs in different cells have access to local CSI and coordinate with each other with limited backhaul information exchange.
While these results are, in spirit, similar to the proposed DBP approach, 
our architecture (i) considers a decentralized architecture in which the computing hardware is \emph{collocated} to support low-latency consensus information exchange, (ii) takes explicit advantage of massive MU-MIMO (the other results in  \cite{dpc_precoding,gs_precoding,admm_mcbf} are for traditional, small-scale MIMO systems), and (iii) proposes a practical way to partition baseband processing that is complementary to CoMP. In fact, one could  integrate DBP together with CoMP to deal with both intra-cell multi-user interference and inter-cell transmission interference more effectively, and to realize decentralized PHY layer processing using our DBP and higher layer (MAC layer, network layer, etc.) resource allocation and coordination with CoMP schemes. In addition, we propose more sophisticated algorithms that enable superior error-rate performance compared  to the methods in \cite{gs_precoding,admm_mcbf}.

\subsubsection{Cloud radio access networks (C-RAN)}  
The idea behind C-RAN is to separate the BS into two modules, a remote radio head (RRH) and a baseband unit (BBU), which are connected via high-bandwidth interconnect. 
The RRHs are placed near the mobile users within the cell, while the BBUs are grouped into a BBU pool for centralized processing and located remotely from RRH~\cite{cloud_arch, cloud_backhaul, cloudran_intro}.
C-RAN and CoMP both coordinate data transmission among multiple cells but with different physical realizations. CoMP integrates each pair of radio heads with associated BBU together and allows low-latency data transfer between each radio head and its corresponding BBU. Different BBUs are separately placed across multiple cells, entailing long latency on coordination among BBUs.  C-RAN, in contrast, shifts the BBU coordination latency in CoMP to the data transfer latency between RRHs and BBUs, since BBUs are now grouped in a pool and can coordinate efficiently. Therefore, whether CoMP or C-RAN is more appropriate depends on whether BBU coordination or RRH-BBU data transfer is more efficient in a real-world deployment. Analogously to CoMP, we could integrate DBP together with C-RAN to exploit the benefits of both technologies. For example, each RRH now can be a large-scale antenna array (requiring higher RRH-BBU interconnection bandwidth).  The associated BBU itself may rely on DBP and perform our algorithms to resolve intra-cell multi-user interference, while coordinating with other BBUs for inter-cell interference mitigation.

\subsubsection{Distributed processing across frequencies}
Existing testbeds, such as the LuMaMi testbed~\cite{lund,lund_lowlatency} and the BigStation~\cite{bigstation}, distribute the necessary baseband processing tasks across frequencies. The idea is to divide the total frequency band into clusters of subcarriers in orthogonal frequency-division multiplexing (OFDM) systems where each frequency cluster is processed concurrently, enabling high degrees of parallelism~\cite{lund,lund_lowlatency, bigstation}.
Unfortunately, each frequency cluster still needs access to all BS antennas, which may result in high interconnect bandwidth. 
Furthermore, the frequency band must somehow be divided either using analog or digital circuitry, and frequency decentralization prevents a straightforward use of other waveform candidates, such as single-carrier frequency-division multiple access (SC-FDMA), filter bank multi-carrier (FBMC), and generalized frequency division multiplexing (GFDM) \cite{tunali2015linear}.
In contrast, our DBP architecture performs decentralization across antennas, which is compatible to most waveforms and requires data transmission only between a subset of antennas and the clusters.
We emphasize, however, that  DBP can be used \emph{together} with frequency decentralization---in fact, our reference GPU  implementation results shown in \fref{sec:GPUcluster} exploit both spatial decentralization and frequency parallelism. 

\subsection{Contributions}
We propose DBP to reduce the raw baseband and chip I/O bandwidths, as well as the signal-processing bottlenecks of massive MU-MIMO systems that perform centralized baseband processing. 
Our main contributions are as follows:
\begin{itemize}
\item We propose DBP, an novel architecture for scalable, TDD-based massive MU-MIMO  BS designs,  which distributes computation across clusters of antennas. 
\item We develop two decentralized algorithms for near-optimal data detection in the massive MU-MIMO uplink; both algorithms trade off error-rate performance vs.\ complexity.  
\item We develop a decentralized beamforming algorithm for the massive MU-MIMO downlink.
\item We perform a simulation-based tradeoff analysis between error-rate performance, consensus data rate, and computational complexity for the proposed decentralized data detection and beamforming algorithms. 
\item We present implementation results for data detection and beamforming on a GPU cluster that showcase the efficacy and scalability of the proposed DBP approach. 
\end{itemize}
Our results demonstrate that DBP enables modular and scalable BS designs for massive MU-MIMO  with thousands of  antenna elements while avoiding excessively high baseband and I/O data rates and significantly reducing the high computational complexity of conventional centralized algorithms.

\subsection{Notation}
Lowercase and uppercase boldface letters designate column vectors and matrices, respectively. 
For a matrix $\bA$, we indicate its transpose and conjugate transpose by  $\bA^T$ and $\bA^H$ respectively. \
The $M\times M$ identity matrix is denoted by $\bI_M$ and the $M\times N$ all-zeros matrix by~$\mathbf{0}_{M\times N}$. 
Sets are denoted by uppercase calligraphic letters; the cardinality of the $\setA$ is denoted by $|\setA|$. 
The real and imaginary parts of a complex scalar $a$ are $\Re\{a\}$ and $\Im\{a\}$, respectively. 
The Kronecker product is $\kron$ and $\Exop[\cdot]$ denotes expectation. 

\subsection{Paper Outline}
The rest of the paper is organized as follows. \fref{sec:arch} details the DBP architecture and introduces the associated uplink and downlink system models. \fref{sec:uplink} proposes two decentralized data detection algorithms. \fref{sec:downlink} proposes a decentralized beamforming algorithm. \fref{sec:results} provides performance and complexity results. \fref{sec:GPUcluster} discusses our  GPU cluster implementations. We conclude in \fref{sec:conclusions}. All proofs are relegated to \fref{app:proofs}.


\section{DBP: Decentralized Baseband Processing}
\label{sec:arch}

We now detail the DBP architecture illustrated in \fref{fig:system_overview} and the system models for the uplink and downlink. 
{We consider a TDD massive MU-MIMO system and we assume a sufficiently long coherence time, i.e., the channel remains constant during both the uplink and downlink phases. In what follows, we focus on narrowband communication; a generalization to wideband systems is straightforward.}

\subsection{Uplink System Model and Architecture}
\label{sec:uplinkarchitecture}
\subsubsection{Uplink system model}
In the uplink phase, $U$ single-antenna\footnote{A generalization to multi-antenna user terminals is straightforward but omitted for the sake of simplicity of exposition.} user terminals communicate with a BS having \mbox{$B\geq U$} antenna elements. Each user encodes its own information bit stream using a forward error correction (FEC) code and maps the resulting coded bit stream to constellation points in the set~$\setO$ (e.g., 16-QAM) using a predefined mapping rule (e.g., Gray mappings). At each user, the resulting constellation symbols are then modulated and transmitted over the wireless channel (subsumed in the ``\textsf{RF}'' block in \fref{fig:system_overview}). The transmit symbols~$s_u$, $u=1,\ldots,U$, of all $U$ users are subsumed in the uplink transmit vector~$\bms^u\in\setO^U$. The baseband-equivalent input-output relation of the (narrowband)  wireless uplink channel is modeled as $\bmy^u=\bH^u\bms^u+\bmn^u$, where $\bmy^u\in\complexset^B$ is the received uplink vector, $\bH^u\in\complexset^{B\times U}$ is the (tall and skinny) uplink channel matrix, and $\bmn^u\in\complexset^B$ is i.i.d.\ circularly-symmetric complex Gaussian noise with variance $\No$ per complex entry. The goal of the BS is to estimate the transmitted code bits given (approximate) knowledge of $\bH^u$ and the received uplink vector $\bmy^u$. This information is then passed to the decoder, which computes estimates for the data bits of each user.

\subsubsection{Decentralized architecture}
Consider the left-hand side (LHS) of \fref{fig:system_overview}. The proposed DBP architecture partitions\footnote{{Other partitioning schemes may be possible but are not considered here.}} the receive vector $\bmy$ into $C$ clusters so that $(\bmy^u)^T=[(\bmy_1^u)^T,\ldots,(\bmy_C^u)^T]$ with $\bmy^u_c\in\complexset^{B_c}$ and $B=\sum_{c=1}^CB_c$. 
As mentioned in \fref{sec:DBPoverview}, we assume clusters of equal size and set $S=B_c$.
By partitioning the uplink channel matrix $(\bH^u)^T=[(\bH^u_1)^T,\ldots,(\bH_c^u)^T]$ \emph{row-wise} into blocks of dimension $\bH^u_c\in\complexset^{B_c\times U}$, $c=1,\ldots,C$, and, analogously, the noise vector as $(\bmn^u)^T=[(\bmn_1^u)^T,\ldots,(\bmn_C^u)^T]$, we can rewrite the uplink input-output relation at each cluster as follows:
\begin{align} \label{eq:uplinkdecomposedmodel}
\bmy_c^u=\bH^u_c\bms^u+\bmn_c^u, \quad c=1,\ldots,C.
\end{align}
The goal of DBP in the uplink is to compute an estimate for $\bms^u$ in a decentralized manner: each cluster $c$ only has access to $\bmy_c^u$,  $\bH^u_c$, and consensus information (see \fref{sec:uplink}). 

As shown in LHS of \fref{fig:system_overview}, each antenna element is associated to local RF processing circuitry; this includes analog and digital filtering, amplification, mixing, modulation, etc. As a consequence, all required digital processing tasks (e.g., used for OFDM processing) are also carried out in a decentralized manner. Even though we consider perfect synchronization and impairment-free transmission (such as carrier frequency and sampling rate offsets, phase noise, or I/Q imbalance), we note that each cluster and the associated RF processing circuitry would be able to separately compensate for such hardware non-idealities with well-established methods~\cite{sync}. 
This key property significantly alleviates the challenges of perfectly synchronizing the clocks and oscillators among the clusters.

\subsubsection{Channel estimation}
During the training phase, each cluster $c$ must acquire local CSI, i.e., compute an estimate of~$\bH^u_c$. To this end, $U$ orthogonal pilots are transmitted from the users prior to the data transmission phase. Since each cluster~$c$ has access to $\bmy_c^u$, it follows from \eqref{eq:uplinkdecomposedmodel} that the associated local channel matrix $\bH^u_c$ can be estimated per cluster. The estimate for the channel matrix (as well as $\bmy_c^u$) is then stored locally at each cluster and \emph{not} made accessible to the other clusters; this prevents a bandwidth-intensive broadcast of CSI (and receive vector data) to all clusters during the training phase. 

\subsubsection{Data detection}
During the data transmission phase, decentralized data detection uses the receive vector $\bmy^u_c$, the associated CSI $\bH^u_c$, and consensus information to generate an estimate of the transmitted data vector $\bms^u$. This estimate is then passed to the decoder which computes estimates for the information bits of each user in a centralized manner; suitable data detection algorithms are proposed in \fref{sec:uplink}.

\subsection{Downlink System Model and  Architecture}

\subsubsection{Downlink system model}
In the downlink phase, the $B$ BS antennas communicate with the $U\leq B$ single-antenna user terminals. 
The information bits for each user are encoded separately using a FEC. The BS then maps the resulting (independent) coded bit streams to constellation points in the alphabet~$\setO$ to form the vector $\bms^d\in\setO^U$.
To mitigate MUI, the BS performs beamforming (BF) {or precoding}, i.e., computes a BF vector $\bmx^d\in\complexset^B$ that is transmitted over the downlink channel. Beamforming requires knowledge of the (short and wide) downlink channel matrix  $\bH^d\in\complexset^{U\times B}$ and the transmit vector $\bms^d\in\setO^U$ to compute a BF vector that satisfies $\bms^d=\bH^d\bmx^d$ (see \fref{sec:downlink} for the details). 
By assuming channel reciprocity, we have the property $\bH^d = (\bH^u)^T$~\cite{mimo_overview, mimo_next_gen}, which implies that the channel matrix estimated in the uplink can be used in the downlink.
The baseband-equivalent input-output relation of the (narrowband)  wireless downlink channel is modeled as $\bmy^d=\bH^d\bmx^d+\bmn^d$, where $\bmy^d\in\complexset^U$ is the receive vector at all users and $\bmn^d\in\complexset^U$ is i.i.d.\ circularly-symmetric complex Gaussian noise with variance $\No$ per complex entry.
By transmitting $\bmx^d$ over the wireless channel, the equivalent input-output relation is given by $\bmy^d=\bms^d+\bmn^d$ and contains no MUI. 
Each of the users then estimates the transmitted code bits from $y^d_u$, $u=1,\ldots,U$. This information is passed to the decoder, which computes estimates for the user's data bits.

\subsubsection{Decentralized architecture}
Consider the right-hand side (RHS) of \fref{fig:system_overview}. Since the partitioning of the BS antennas was fixed for the uplink (cf.~\fref{sec:uplinkarchitecture}), the BF vector~$\bmx^d$ must be partitioned into $C$ clusters so that $(\bmx^d)^T=[(\bmx_1^d)^T,\ldots,(\bmx_C^d)^T]$ with $\bmx^d_c\in\complexset^{B_c}$.
By using reciprocity and the given antenna partitioning, each cluster~$c$ has access to only $\bH^d_c = (\bH^u_c)^T$. 
With this partitioning, we can rewrite the downlink input-output relation as follows:
\begin{align} \label{eq:downlinkdecomposedmodel}
\bmy^d= \textstyle \sum_{c=1}^C\bH^d_c\bmx^d_c+\bmn^d.
\end{align}
The goal of DBP in the downlink is to compute all local BF vectors $\bmx^d_c$, $c=1,\ldots,C$, in a decentralized manner: each cluster $c$  has access to only $\bms^d$,  $\bH^d_c$, and consensus information (see \fref{sec:downlink} for more details). 

As shown in the RHS of \fref{fig:system_overview}, each antenna element is associated to local RF processing circuitry. Analogously to the uplink, the required analog and digital signal processing tasks (e.g., used for OFDM modulation or impairment compensation) can be carried out in a decentralized manner, which alleviates the challenges of perfectly synchronizing the clusters.

\subsubsection{Beamforming}
In the downlink phase, decentralized BF uses the transmit vector $\bms^d$, decentralized CSI $\bH^d_c$, and consensus information in order to generate BF vectors $\bmx^d_c$ that satisfy $\bms^d=\sum_{c=1}^C\bH^d_c\bmx^d_c$. This ensures that transmission of the vectors~$\bmx_c^d$ removes MUI; a suitable algorithm is detailed in \fref{sec:downlink}.


\section{Decentralized Uplink: Data Detection}
\label{sec:uplink}

We now propose two decentralized data detection algorithms for the massive MU-MIMO uplink.  We start by discussing the general equalization problem and then, detail our novel ADMM and CG-based data detection algorithms. 
To simplify notation, we omit the uplink superscript $^u$ in this section.

\subsection{Equalization-Based Data Detection}

In order to arrive at computationally efficient algorithms for decentralized data detection, we focus on equalization-based methods. Such methods  contain an equalization stage and a detection stage. For the equalization stage, we are interested in solving the following equalization problem
\begin{align*}
\text{(E0)} \quad  \hat\bmx = \argmin_{\bms\in\complexset^U} \,  g(\vecs) + \textstyle \frac{1}{2}\vecnorm{\bmy-\bH\bms}_2^2
\end{align*}
in a decentralized manner. Here, the function $g(\vecs)$ is a convex (but not necessarily smooth or bounded) regularizer, which will be discussed in detail below. 
For the detection stage, the result~$\hat\bmx$ of the equalization problem~(E0) can either be  sliced entry-wise to the nearest constellation point in $\setO$ to perform hard-output data detection or used to compute approximate soft-output values e.g., log-likelihood ratio (LLR) values~\cite{asic_det}. 

\revision{For zero-forcing (ZF) and minimum mean-squared error (MMSE) data detection, we set the regularizer to $g(\vecs)=0$ and  $g(\vecs)={\No/}{(2\Es)}\vecnorm{\vecs}^2_2$, respectively, where $\Es=\Exop[|s_u|^2]$ for $u=1,\ldots,U$, is the expected per-user transmit energy.}\footnote{For the sake of simplicity, we assume an equal transmit power at each user. An extension to the general case is straightforward.}
The generality of the equalization problem (E0) also encompasses more powerful data detection algorithms. In particular, we can set $g(\bms)=\chi(\bms\in\setC)$, where $\chi(\bms\in\setC)$ is the characteristic function that is zero if $\bms$ is in some convex set $\setC$ and infinity otherwise. 
Specifically, to design data-detection algorithms that outperform ZF or MMSE data detection, we can use the convex polytope around the constellation set~$\setO$, which is given by 
\begin{align*}
\textstyle  \setC = \left\{\sum_{i=1}^{\abs{\setO}}\alpha_i s_i \mid (\alpha_i\geq0,\forall i) \wedge \sum_{i=1}^{\abs{\setO}}\alpha_i=1 \right\}.
\end{align*}
For QPSK with $\setO=\{\pm1\pm i\}$, the convex set $\setC$ is simply a box with radius $1$ (i.e., side length of 2) centered at the origin. In this case, (E0) corresponds to the so-called box-constrained equalizer~\cite{box} which was shown recently to (often significantly) outperform linear ZF or MMSE data detection~\cite{box3,jeon2016performance}. In addition, box-constrained equalization does not require knowledge of the noise variance~$\No$, which is in stark contrast to the traditional MMSE equalizer. The decentralized equalization algorithm proposed next enables the use of such powerful regularizers.

\subsection{Decentralized Equalization via ADMM}
\label{sec:ADMMequalization}

To solve the equalization problem (E0) in a decentralized fashion, we make use of the ADMM framework~\cite{admm}.
We first introduce~$C$ auxiliary variables $\vecz_c=\vecs$, $c=1,\ldots,C$, which allow us to rewrite~(E0) in the equivalent form  
\begin{align*}
\text{(E0$'$)} \quad  \hat\bmx =\!\! \argmin_{\bms\in\complexset^U,\, \vecz_c=\vecs,\,  c=1,\ldots,C} \, g(\vecs) + \textstyle \sum_{c=1}^C\frac{1}{2}\vecnorm{\bmy_c-\bH_c\bmz_c}_2^2.
\end{align*}
Note that the added constraints in \text{(E0$'$)} enforce that each local vector $\vecz_c$ agrees with the global value of $\vecs.$ 
As detailed in~\cite{admm}, these constraints can be enforced by introducing Lagrange multipliers $\{\bmlambda_c\}_{c=1}^C$ for each cluster, and then computing a saddle point (where the augmented Lagrangian is minimal for~$\bms$ and $\bmz$, and maximal for $\bmlambda$) of the so-called scaled augmented Lagrangian function, which is defined as
\begin{align*} 
& \setL(\bms,\bmz,\bmlambda) = g(\bms) \\
& \quad \qquad + \textstyle \sum_{c=1}^C\Big\{\frac{1}{2}\vecnorm{\bmy_c-\bH_c\bmz_c}_2^2 + \frac{\rho}{2}\vecnorm{\bms-\bmz_c-\bmlambda_c}_2^2\Big\} 
\end{align*}
\!\!for some fixed penalty parameter $\rho>0$. Here, we stack all $C$ auxiliary variables into the vector $\bmz^T=[\bmz_1^T\cdots\,\bmz_C^T]$ and stack all $C$  scaled Lagrange multiplier vectors into the vector $\bmlambda^T=[\bmlambda_1^T\cdots\,\bmlambda_C^T]$, where $\bmz_c,\bmlambda_c\in\complexset^U$. 

The saddle-point formulation of (E0$'$) is an example of a global variable consensus problem~\cite[Sec.~7.1]{admm} and can be solved using ADMM. 
We initialize $\bms^{(1)}=\bZero_{U \times 1}$ and \mbox{$\bmlambda_c^{(1)}=\bZero_{U \times 1}$} for $c=1,\ldots,C$, and carry out the following iterative steps:
\begin{align*}
\text{(E1)} \!\!\!\!  &&  \bmz_c^{(t+1)} & = \argmin_{\bmz_c\in\complexset^U}   \textstyle \frac{1}{2}\vecnorm{\bmy_c\!-\!\bH_c\bmz_c}_2^2 + \frac{\rho}{2} \big\|\bms^{(t)}\!-\!\bmz_c\!-\!\bmlambda_c^{(t)}\big\|_2^2 \\
\text{(E2)} \!\!\!\!  && \bms^{(t+1)} & = \argmin_{\bms\in\complexset^U}   \textstyle g(\bms) + \sum_{c=1}^C {\frac{\rho}{2}} \big\|\bms-\bmz_c^{(t+1)}-\bmlambda_c^{(t)}\big\|_2^2 \\
\text{(E3)} \!\!\!\!  && \bmlambda_c^{(t+1)} & = \bmlambda_c^{(t)} - \gamma\big(\bms^{(t+1)}-\bmz_c^{(t+1)}\big)
\end{align*}
for the iterations $t=1,2,\ldots$ until convergence or a maximum number $T_\text{max}$ of iterations has been reached.
The parameter $\rho>0$ controls the step size and $\gamma=1$ is a typical choice that guarantees convergence. 
See~\cite{admm_converge1} for a more detailed discussion on the convergence of ADMM. 

{Steps (E1) and (E2) can be carried out in a decentralized and parallel manner, i.e., each cluster $c=1,\ldots,C$ only requires access to local variables and local  channel state information, as well as the consensus vectors $\bms^{(t)}$ and $\bms^{(t+1)}$.}
Step~(E2) updates the consensus vector.  While the vectors $\{\bmz_c^{(t+1)}\}$ and $\{\bmlambda_c^{(t)}\}$ for every cluster appear in (E2), it is known that this can be computed using only the global average of these vectors, which is easily stored on a fusion node \cite{admm}. The architecture proposed in \fref{sec:arch} can compute these averages and perform this update in an efficient manner.
We next discuss the key details of the proposed decentralized data detection algorithm. 

\subsection{ADMM Algorithm Details and Decentralization}

\subsubsection{Step (E1)}
This step corresponds to a least-squares (LS) problem that can be solved in closed form and independently on each cluster. For a given cluster $c$, we can rewrite the minimization in Step (E1) in more compact form as
\begin{align*}
 \bmz_c^{(t+1)} & = \argmin_{\bmz_c\in\complexset^U} \,  
\vecnorm{
\!\left[\begin{array}{c}
\bmy_c \\
\sqrt{\rho}(\bms^{(t)}-\bmlambda_c^{(t)})
\end{array}\right]\!
-
\!\left[\begin{array}{c}
\bH_c\\
\sqrt{\rho}\,\bI_U
\end{array}\right]\!
\bmz_c}_2^2, 
\end{align*}
which  has the following closed-form solution:
\begin{align} \label{eq:StepE1variant1}
 \bmz_c^{(t+1)} & =\bmy_c^\text{reg} + \rho\bB^{-1}_c(\bms^{(t)}-\bmlambda_c^{(t)}).
\end{align}
Here, $\bmy_c^\text{reg}=\bB_c^{-1}\bH_c^H\bmy_c$ is the regularized estimate with $\bB_c^{-1}=(\bH_c^H\bH_c+\rho\bI_{U})^{-1}$. To reduce the amount of recurrent computations, we can precompute $\bB_c^{-1}$ and reuse the result in each iteration. 
For situations where the cluster size $S$ is smaller than the number of users $U$, we can use the Woodbury matrix identity~\cite{woodbury} to derive the  following equivalent update:
\begin{align} \label{eq:StepE1variant2}
 \bmz_c^{(t+1)} & =\bmy_c^\text{reg} + (\bI_U -\bH_c^H\bA^{-1}_c\bH_c)(\bms^{(t)}-\bmlambda_c^{(t)}).
\end{align}
Here, $\bmy_c^\text{reg}=\bH_c^H\bA_c^{-1}\bmy_c$ is a regularized estimate of the transmit vector with $\bA_c^{-1}=(\bH_c\bH_c^H+\rho\bI_{S})^{-1}$. This requires the inversion of an $S\times S$ matrix, which is more easily computed than  the $U\times U$ inverse required by~\eqref{eq:StepE1variant1}. 
{We note that whether~\eqref{eq:StepE1variant1} or~\eqref{eq:StepE1variant2} leads to lower computational complexity depends on $U$, $S$, and the number of ADMM iterations (see \fref{sec:compl}).}

\begin{algorithm}[t]
\caption{Decentralized ADMM-based Data Detection} 
\label{alg:admm_det}
\begin{algorithmic}[1]
\small

\STATE \textbf{Input}: $\mathbf{y}_c$, $\mathbf{H}_c$, $c=1,2,\ldots,C$, $\rho$, $\gamma$, $\No$, and $\Es$
\STATE \textbf{\em Preprocessing}:
\IF{$S \leq U$}  
\STATE \revision{$\mathbf{A}_c^{-1}=(\mathbf{H}_c\mathbf{H}_c^H+\rho\mathbf{I}_{S})^{-1}$}
\STATE $\mathbf{y}_c^\text{reg}=\mathbf{H}_c^H\mathbf{A}_c^{-1}\mathbf{y}_c$

\ELSE
\STATE \revision{$\mathbf{B}_c^{-1}=(\mathbf{H}_c^H\mathbf{H}_c+\rho\mathbf{I}_{U})^{-1}$}
\STATE $\mathbf{y}_c^\text{reg}=\mathbf{B}_c^{-1}\mathbf{H}_c^H\mathbf{y}_c$

\ENDIF

\STATE \textbf{\em ADMM iterations}:
\STATE \textbf{Init}: $\boldsymbol\lambda_c^{(1)}=\mathbf{0},\,\mathbf{z}_c^{(1)}=\mathbf{y}_c^\text{reg},\,\mathbf{s}^{(1)}=\big({\frac{\No}{\rho\Es}}+C\big)^{-1}(\sum_{c=1}^C\mathbf{z}_c^{(1)})$ 

\FOR{$t=2,3,\ldots,T_\text{max}$} 
\STATE $\boldsymbol\lambda_c^{(t)}=\boldsymbol\lambda_c^{(t-1)}+\gamma(\mathbf{z}^{(t-1)}_c-\mathbf{s}^{(t-1)})$
\IF{$S \leq U$} 
\STATE $\mathbf{z}^{(t)}_c\!=\mathbf{y}_c^\text{reg}+(\bms^{(t-1)}-\boldsymbol\lambda^{(t)}_c)-\mathbf{H}_c^H\mathbf{A}_c^{-1}\mathbf{H}_c(\bms^{(t-1)}-\boldsymbol\lambda^{(t)}_c)$
\ELSE
\STATE $\mathbf{z}_c^{(t)}\!=\mathbf{y}_c^\text{reg}+\rho\mathbf{B}^{-1}_c(\bms^{(t-1)}-\boldsymbol\lambda^{(t)}_c)$
\ENDIF
\STATE $\mathbf{w}^{(t)}_c=\mathbf{z}_c^{(t)}+\boldsymbol\lambda^{(t)}_c$
\STATE $\mathbf{w}^{(t)}=\sum_{c=1}^C\mathbf{w}^{(t)}_c${\qquad\qquad} // Consensus  
\STATE $\mathbf{s}^{(t)}=\big({\frac{\No}{\rho\Es}}+C\big)^{-1}\mathbf{w}^{(t)}$ {\quad} 

\ENDFOR

\STATE \textbf{Output}: $\hat{\bmx}=\bms^{(T_\text{max})}$
\end{algorithmic}
\end{algorithm}

\subsubsection{Step (E2)}
This step requires gathering of local computation results, averaging the sum in a centralized manner, and distributing the averaged consensus information. To reduce the amount of data that must be exchanged, each cluster only communicates the intermediate variable $\mathbf{w}_c^{(t)}=\bmz_c^{(t+1)}+\bmlambda_c^{(t)},$ and only the average of these vectors is used on the fusion node. This simplification is accomplished using the following lemma; a proof is given in \fref{app:D2simplification}. 
 
\begin{lem} \label{lem:D2simplification}
The problem in Step (E2) simplifies to
\begin{align} \label{eq:D2averaging}
\bms^{(t+1)}  = \argmin_{\bms\in\complexset^U}  \, g(\bms) + \textstyle {\frac{C\rho}{2}}\vecnorm{\bms-\bmv^{(t)}}_2^2
\end{align}
with $\bmv^{(t)}=\frac{1}{C}\bmw^{(t)}=\frac{1}{C} \sum_{c=1}^C\bmw_c^{(t)}$ and $\bmw_c^{(t)}=\bmz_c^{(t+1)}+\bmlambda_c^{(t)}.$
\end{lem}

Computation of \eqref{eq:D2averaging} requires two parts. The first part corresponds to a simple averaging procedure to obtain $\bmv^{(t)}$, which can be carried out via sum reduction in a tree-like fashion followed by centralized averaging. 
The second part is the minimization in \eqref{eq:D2averaging} that is known as the proximal operator for the function $g(\bms)$~\cite{convex}. 
For ZF, MMSE, and box-constrained equalization with QAM alphabets, the proximal operator has the following simple closed-form expressions:
\begin{align*}
\text{(E2-ZF)} && \bms^{(t+1)} & = \bmv^{(t)} & \\
\text{(E2-MMSE)} && \bms^{(t+1)} & = \textstyle {\frac{C\rho\Es}{\No+C\rho\Es}\bmv^{(t)}}  \\
\text{(E2-BOX)} && s^{(t+1)}_u & = \sign(\Re\{v_u^{(t)}\}) \min\{|\Re\{v_u^{(t)}\}|,r\} \\
 && & \quad + i\sign(\Im\{v_u^{(t)}\}) \min\{|\Im\{v_u^{(t)}\}|,r\}
\end{align*}
for $u=1,\ldots,U$.
Here, (E2-BOX) is the orthogonal projection of the vector $\bmv^{(t)}$ onto the hypercube with radius $r$ that covers the QAM constellation. 
{For BPSK, the proximal operator corresponds to the orthogonal projection onto the real line between $[-r,+r]$ and is given by $s^{(t+1)}_u  = \sign(\Re\{v_u^{(t)}\}) \min\{|\Re\{v_u^{(t)}\}|,r\}$, $u=1,\ldots,U$.}

After computation of \eqref{eq:D2averaging}, the consensus vector $\bms^{(t+1)}$ needs to be distributed to all $C$ clusters. 
In practice, we distribute~$\bmw^{(t)}$ as soon as it is available, and the scaling steps to get $\bms^{(t+1)}$ from~$\bmw^{(t)}$ are computed locally on each cluster after it receives~$\bmw^{(t)}.$  With this approach, no cluster waits for the computation of $\bms^{(t+1)}$ on a central/master worker (fusion node) before ADMM iterations proceed.

\subsubsection{Step (E3)}
This step can be carried out independently in each cluster after $\bms^{(t+1)}$ has been calculated.

We summarize the resulting decentralized  ADMM procedure for MMSE equalization in Algorithm~\ref{alg:admm_det}. 
{The equalization output is the consensus vector $\hat\bmx=\mathbf{s}^{(T_\text{max})}$. Note that Algorithm~\ref{alg:admm_det} slightly deviates from the procedure outlined in \fref{sec:ADMMequalization}. Specifically, Algorithm 1 performs the steps in the following order: Step (E3), Step (E1), and Step (E2); this is due to the fact that the global equalizer output $\hat\bmx$ results from Step (E2). We note that re-ordering the steps as in Algorithm 1 has no effect on the convergence and the equalization result.}

We will analyze the algorithm's complexity\footnote{The operataion $\mathbf{H}_c^H\mathbf{A}_c^{-1}\mathbf{H}_c$ on line 15 of Algorithm~\ref{alg:admm_det} could be computed once in a preprocessing stage to avoid recurrent computations during the iterations. Instead,  in Algorithm~\ref{alg:admm_det} we directly compute $\mathbf{H}_c^H\mathbf{A}_c^{-1}\mathbf{H}_c(\bms^{(t-1)}-\boldsymbol\lambda^{(t)}_c)$ in each iteration because this approach requires only three matrix-vector multiplications per iteration; precomputing $\mathbf{H}_c^H\mathbf{A}_c^{-1}\mathbf{H}_c$ requires two costly matrix-matrix multiplications. Hence, our complexity analysis in \fref{sec:compl} refers to  the procedure detailed in Algorithm~\ref{alg:admm_det}.} in \fref{sec:compl}; a GPU cluster implementation will be discussed in \fref{sec:GPUcluster}.

\subsection{Decentralized Equalization via Conjugate Gradients}
\label{sec:dcg_det}
If the regularization function $g(\bms)$ of (E0) is quadratic, as in the case for MMSE equalization where \revision{$g(\bms)={\No/}{(2\Es)}\|\bms\|_2^2$}, then we can solve (E0) with an efficient decentralized conjugate gradient (CG) method~\cite{cg_mimo,cg_gpu,goldstein2010high}.  
Our method builds on the CG algorithm used in~\cite{cg_mimo} for \emph{centralized} equalization-based data detection in massive MU-MIMO systems. 
Our  idea is to  break all centralized computations that rely on global CSI and receive data (i.e., $\mathbf{H}$ and $\mathbf{y}$) into smaller, independent problems that only require local CSI and receive data ($\mathbf{H}_c$ and~$\mathbf{y}_c$).
The centralized CG-based detector in~\cite{cg_mimo} involves two stages: a preprocessing stage for calculating the MRC output~$\mathbf{y}^{\text{MRC}}$ and a CG iteration stage to estimate $\hat{\mathbf{x}}$.

In the preprocessing stage, we rewrite the MRC vector $\mathbf{y}^{\text{MRC}}=\mathbf{H}^H\mathbf{y}$ as $\mathbf{y}^\text{MRC}=\sum_{c=1}^C \mathbf{H}_c^H\mathbf{y}_c$, which decentralizes the preprocessing stage. Specifically, each cluster computes $\mathbf{H}_c^H\mathbf{y}_c$; the results of each cluster are then summed up in a centralized manner to obtain the MRC output $\mathbf{y}^{\text{MRC}}$.

For the CG iteration stage, we need to update the estimated transmit vector and a number of intermediate vectors required by the CG algorithm  (see~\cite{cg_mimo} for the details). While most operations are not directly dependent on global CSI $\mathbf{H}$ but on intermediate results, the update of the following vector
\begin{align} \label{eq:CGiteration}
\mathbf{e}^{(t)}=\textstyle \big(\rho\mathbf{I} + \mathbf{H}^H\mathbf{H}\big)\mathbf{p}^{(t-1)},
\end{align} 
requires direct access to the global channel matrix $\mathbf{H}$ and thus, must be decentralized. Here, $\rho={\No/}{\Es}$ for MMSE equalization and $\rho=0$ for zero-forcing equalization. 
It is key to realize that the Gram matrix can be written as $\mathbf{H}^H\mathbf{H}=\sum_{c=1}^C \mathbf{H}_c^H\mathbf{H}_c $. Hence, we  can reformulate \eqref{eq:CGiteration} as 
\begin{align}  \label{eq:CGdecentralization}
\mathbf{e}^{(t)}= \textstyle \rho\mathbf{p}^{(t-1)} + \sum_{c=1}^C \mathbf{H}_c^H\mathbf{H}_c\mathbf{p}^{(t-1)}.
\end{align}
Put simply, by locally computing $\mathbf{w}_c^{(t)}=\mathbf{H}_c^H\mathbf{H}_c\mathbf{p}^{(t-1)}$ at each antenna cluster, we can obtain the result in \eqref{eq:CGdecentralization} by performing the following centralized computations that do not require global CSI: $\mathbf{w}^{(t)}=\sum_{c=1}^C \mathbf{w}_c^{(t)}$ and $\mathbf{e}^{(t)}=\rho\mathbf{p}^{(t-1)}+\mathbf{w}^{(t)}$. 

\begin{algorithm}[t]
\caption{Decentralized CG-based Data Detection \label{alg:CG}}
\small
\begin{algorithmic}[1]
\STATE \textbf{Input:} $\mathbf{H}_c$, $c=1,\ldots,C$, and $\mathbf{y}_c$, and $\rho$
\STATE \textbf{\em Preprocessing:}

\STATE \quad $\mathbf{y}_c^\text{MRC}=\mathbf{H}_c^H\mathbf{y}_c$ \,\qquad \quad // Decentralized 
\STATE \quad $\mathbf{y}^\text{MRC}=\sum_{c=1}^C \mathbf{y}_c^\text{MRC}$ \quad // Centralized 

\STATE \textbf{\em CG iterations:}
\STATE \textbf{Init:} $\mathbf{r}^{(0)}=\mathbf{y}^\text{MRC}, \mathbf{p}^{(0)}=\mathbf{r}^{(0)}, \mathbf{x}^{(0)}=\mathbf{0}$

\FOR{$t = 1,\ldots, T_\text{max}$}
\STATE $\text{Decentralized (each cluster $c$ performs the same operation):}$ 
\STATE $\quad\mathbf{w}_c^{(t)}=\mathbf{H}^H_c\mathbf{H}_c\mathbf{p}^{(t-1)}$
\STATE  $\text{Centralized (consensus on a centralized processing unit):} $
\STATE $\quad\mathbf{w}^{(t)}=\sum_{c=1}^c \mathbf{w}_c^{(t)}$ \quad // Consensus 

\STATE $\text{Decentralized   (each cluster $c$ performs the same operations):}$
\STATE $\quad\mathbf{e}^{(t)}=\rho\mathbf{p}^{(t-1)}+\mathbf{w}^{(t)}$
\STATE $\quad\alpha=\|\mathbf{r}^{(t-1)}\|^2/((\mathbf{p}^{(t-1)})^H\mathbf{e}^{(t)})$
\STATE $\quad\mathbf{x}^{(t)}=\mathbf{x}^{(t-1)}+\alpha\mathbf{p}^{(t-1)}$
\STATE $\quad\mathbf{r}^{(t)}=\mathbf{r}^{(t-1)}-\alpha\mathbf{e}^{(t-1)}$
\STATE $\quad\beta=\|\mathbf{r}^{(t)}\|^2/\|\mathbf{r}^{(t-1)}\|^2$
\STATE $\quad\mathbf{p}^{(t)}=\mathbf{r}^{(t)}+\beta\mathbf{p}^{(t-1)}$
\ENDFOR
\STATE \textbf{Output:} $\hat{\mathbf{x}}=\mathbf{x}^{(T_\text{max})}$
\end{algorithmic}
\end{algorithm}

The decentralized CG-based data detection algorithm is summarized in Algorithm~\ref{alg:CG}.
The computations of $\mathbf{e}^{(t)}$, $\mathbf{x}^{(t)}$, $\mathbf{r}^{(t)}$, and $\mathbf{p}^{(t)}$ do not require access to the (global) channel matrix~$\mathbf{H}$ and can be carried out in a centralized processing unit. We must, however, broadcast the vector $\mathbf{p}^{(t)}$ to each antenna cluster before the decentralized update of $\mathbf{w}_c^{(t+1)}$ in the next iteration can take place. Alternatively, we can directly broadcast the consensus vector $\mathbf{w}^{(t)}$, so that each antenna cluster can simultaneously compute their \emph{own copy} of $\mathbf{e}^{(t)}$, $\mathbf{x}^{(t)}$, $\mathbf{r}^{(t)}$, and $\mathbf{p}^{(t)}$ in a decentralized manner to ensure the local existence of $\mathbf{p}^{(t)}$ for updating $\mathbf{w}_c^{(t+1)}$. {With  this alternative approach, we can completely shift the complexity from the centralized processing unit to each cluster, leaving the calculation of  $\mathbf{w}^{(t)}$ as the only centralized computation in a CG iteration. This approach also enables the concatenation of data gathering and broadcasting, which can be implemented using a single message-passing function (see \fref{sec:GPUcluster} for the details).}\footnote{The Gram matrix $\mathbf{G}_c=\mathbf{H}_c^H\mathbf{H}_c$ can be precomputed to avoid recurrent computations (line 9 in Algorithm~\ref{alg:CG}). However, practical systems only need a small number of CG iterations, and $\mathbf{H}_c^H\mathbf{H}_c\mathbf{p}^{(t-1)}$ at line 9 is computed using two matrix-vector multiplications, which avoids the expensive matrix-matrix multiplication needed to form~$\mathbf{G}_c.$ }


\section{Decentralized Downlink: Beamforming}
\label{sec:downlink}

We now develop a decentralized beamforming algorithm for the massive MU-MIMO downlink. We start by discussing the general beamforming (or precoding) problem, and then detail our ADMM-based beamforming algorithm. To simplify notation, we omit the downlink superscript $^d$. 

\subsection{Beamforming Problem}

We solve the following beamforming problem
\begin{align*}
\text{(P0)} \quad  \hat\bmx = \argmin_{\bmx\in\complexset^B} \, \vecnorm{\bmx}_2 \quad \text{subject to}\, \vecnorm{\bms-\bH\bmx}_2\leq \varepsilon,
\end{align*}
which aims at minimizing the instantaneous transmit energy while satisfying the precoding constraint $\vecnorm{\bms-\bH\bmx}_2\leq \varepsilon$. By defining the residual interference as $\vece=\bms-\bH\hat\bmx$, we see that transmission of the solution vector $\hat\bmx$ of (P0) leads to the input-output relation $\vecy=\vecs+\vece+\vecn$ with $\vecnorm{\vece}_2\leq\varepsilon$. Hence, each user only sees their dedicated signal contaminated with  Gaussian noise $\vecn$ and residual interference $\vece$, whose energy can be controlled by the parameter  $\varepsilon\geq0$. 
By setting \mbox{$\varepsilon=0$}, this problem has a well-known closed-form solution and corresponds to the so-called zero-forcing (ZF) beamformer, which is given by $\hat\vecx=\bH^H(\bH\bH^H)^{-1}\bms$ assuming that $U\leq B$ and $\bH$ is full rank. 
Our goal is to develop an algorithm that computes the solution of (P0) in a decentralized fashion.  

\subsection{Decentralized Beamforming via ADMM}
\label{sec:admm_prec}
%
By introducing $C$ auxiliary variables $\vecz_c=\bH_c\vecx_c$, $c=1,\ldots,C$, we can rewrite (P0) in the following equivalent form:
\begin{align*}
\text{(P0$'$)} \quad  \hat\bmx =   \argmin_{\bmx\in\complexset^B} \, \vecnorm{\vecx}_2 \quad 
& \text{subject to } \textstyle  \big\|{\bms-\sum_{c=1}^C\vecz_c}\big\|_2\leq\varepsilon  \\
& \text{and } \vecz_c=\bH_c\vecx_c, \,\, c=1,\ldots,C.
\end{align*}
Here, $\bH_c$ is the downlink channel matrix at cluster $c$. The solution to the beamforming problem (P0$'$) corresponds to
a saddle point of the scaled augmented Lagrangian function:
\begin{align*}
\setL(\bms,\bmz,\bmlambda) = \textstyle \frac{1}{2}\vecnorm{\vecx}_2^2 +  \sum_{c=1}^C\frac{\rho}{2}\vecnorm{\bH_c\bmx_c-\bmz_c-\bmlambda_c}_2^2  + \setX\!\left(\vecz\right),
\end{align*}
where $\setX(\vecz)$ is the characteristic function for the convex constraint of the beamforming problem (P0), i.e., $\setX(\vecz)=0$ if $\|\bms-\sum_{c=1}^C\bmz_c\|_2\leq\varepsilon$ and $\setX(\vecz)=\infty$ otherwise. The problem (P0) corresponds to a sharing consensus problem with regularization~\cite[Sec.~7.3]{admm}. 

In order to arrive at a decentralized beamforming algorithm, we now use the ADMM framework to find a solution to (P0$'$). We initialize\footnote{\revision{This initializer is a properly-scaled version of the MRC beamforming vector and exhibits good performance for the considered channel model.}}  $\bmz_c^{(1)}=\max\{U/B,1/C\}\bms$. 
We then perform the following three-step procedure until convergence or a maximum number of iterations has been reached: 
\begin{align*}
\text{(P1)}&\,\,\,\, \bmx_c^{(t+1)} \!=\! \argmin_{\bmx_c\in\complexset^S} \textstyle \frac{1}{2}\vecnorm{\vecx_c}_2^2+ \frac{\rho}{2}\big\|{\bH_c\bmx_c-\bmz_c^{(t)}-\bmlambda^{(t)}_c}\big\|_2^2\\
\text{(P2)}&\,\,\,\,\bmz^{(t+1)}  \!=\! \argmin_{\bmz_c\in\complexset^U} \textstyle \!\sum\limits_{c=1}^C \frac{\rho}{2} \big\|{\bH_c\bmx_c^{(t+1)}\!-\!\bmz_c \!-\!\bmlambda_c^{(t)}}\big\|_2^2\!+\!\setX\!\left(\vecz\right)\\
\text{(P3)}&\,\,\,\,\bmlambda^{(t+1)}_c  \!=\!  \bmlambda^{(t)}_c  - \gamma\big(\bH_c\bmx_c^{(t+1)}- \bmz_c^{(t+1)}\big).
\end{align*}
Here, $\bmz_c$ is the local beamforming output and $\bmz$ is the consensus solution of (P2). The parameter $\rho>0$ affects the step size and $\gamma=1$ ensures convergence of the algorithm. 
While both the Steps~(P1) and~(P3) can efficiently be computed in a decentralized manner, it is not obvious how Step~(P2) can be decentralized. We next show the details to transform Step~(P2) into a form that requires simple global averaging.

\subsection{ADMM Algorithm Details and Decentralization}

\subsubsection{Step (P1)}
Analogous to Step (E1), this step corresponds to a LS problem that can be solved in closed form and independently in every cluster. For a given cluster $c=1,\ldots,C$, we can rewrite the minimization in (P1) as 
\begin{align*}
 \bmx_c^{(t+1)} & = \argmin_{\bmx_c\in\complexset^S} \,  
\vecnorm{
\!\left[\begin{array}{c}
\sqrt{\rho}(\bmz^{(t)}_c+\bmlambda_c^{(t)}) \\ 
\bZero_{S \times 1} \\
\end{array}\right]\!
-
\!\left[\begin{array}{c}
\sqrt{\rho}\bH_c\\
\bI_S
\end{array}\right]\!
\bmx_c}_2^2, 
\end{align*}
which  has the following closed-form solution:
\begin{align*}
 \bmx_c^{(t+1)} & = \bA^{-1}_c\bH_c^H(\bmz^{(t)}_c+\bmlambda_c^{(t)}).
\end{align*}
%
%
Here, $\bA_c^{-1}=(\bH_c^H\bH_c+\rho^{-1}\bI_{S})^{-1}$ requires the computation of an $S\times S$ matrix inverse. 
If the cluster size $S$ is larger than the number of users $U$, then we can use the Woodbury matrix identity~\cite{woodbury}
to derive the following equivalent update:
\begin{align*}
 \bmx_c^{(t+1)} & = \bH_c^H\bB^{-1}_c(\bmz^{(t)}_c+\bmlambda_c^{(t)}).
\end{align*}
Here,  $\bB_c^{-1}=(\bH_c\bH_c^H+\rho^{-1}\bI_{U})^{-1}$ requires the computation of an $U\times U$ matrix inverse. 
We note that $U$, $S$, and the number of iterations can determine which of the two $\bmx_c^{(t+1)}$ variations leads to lower overall computational complexity. 

\begin{algorithm}[t]
\caption{Decentralized ADMM-based Beamforming}
\label{alg:admm_prec}
\begin{algorithmic}[1]
\small

\STATE \textbf{Input:} $\mathbf{s}$, $\mathbf{H}_c$, $c=1,2,\ldots,C$, $\rho$, and $\gamma$

\STATE \textbf{\emph{Preprocessing:}}
\IF{$S \leq U$}  
\STATE $\mathbf{A}_c^{-1}=(\mathbf{H}_c^H\mathbf{H}_c+\rho^{-1}\mathbf{I}_{S})^{-1}$
\ELSE
\STATE $\mathbf{B}_c^{-1}=(\mathbf{H}_c\mathbf{H}_c^H+\rho^{-1}\mathbf{I}_{U})^{-1}$
\ENDIF

\STATE \textbf{\emph{ADMM iterations:}}
\STATE \textbf{Init:} $\mathbf{z}_c^{(1)}=\max\{U/B,1/C\}\mathbf{s},\,\boldsymbol\lambda_c^{(1)}=\mathbf{0}$
\STATE \quad\quad $\mathbf{x}_c^{(1)}=\mathbf{A}_c^{-1}\mathbf{H}_c^H\mathbf{z}_c^{(1)} (S\leq U)$ or $\mathbf{H}_c^H\mathbf{B}_c^{-1}\mathbf{z}_c^{(1)} (S>U)$

\FOR{$t=2,3,\ldots,T_\text{max}$} 
\STATE $\mathbf{m}^{(t-1)}_c=\mathbf{H}_c\mathbf{x}^{(t-1)}_c$
\STATE $\mathbf{w}^{(t-1)}_c=\mathbf{m}^{(t-1)}_c-\boldsymbol\lambda_c^{(t-1)}$
\STATE $\mathbf{w}^{(t-1)}=\sum_{c=1}^C\mathbf{w}^{(t-1)}_c${\qquad} // Consensus  
\STATE  $\bmz^{(t)}_c = \bmw^{(t-1)}_c + C^{-1} (\bms-\bmw^{(t-1)})$
\STATE $\boldsymbol\lambda^{(t)}_c=\boldsymbol\lambda^{(t-1)}_c  - \gamma(\mathbf{m}^{(t-1)}_c- \mathbf{z}_c^{(t)})$

\IF{$S \leq U$}  
\STATE $\mathbf{x}_c^{(t)}=\mathbf{A}_c^{-1}\mathbf{H}_c^H(\mathbf{z}_c^{(t)}+\boldsymbol\lambda_c^{(t)})$
\ELSE
\STATE $\mathbf{x}_c^{(t)}=\mathbf{H}_c^H\mathbf{B}_c^{-1}(\mathbf{z}_c^{(t)}+\boldsymbol\lambda_c^{(t)})$
\ENDIF

\ENDFOR

\STATE \textbf{Output:} $\mathbf{\hat{x}}=[\mathbf{x}^{(T_\text{max})}_1;\mathbf{x}^{(T_\text{max})}_2;\cdots;\mathbf{x}^{(T_\text{max})}_C]$
\end{algorithmic}
\end{algorithm}

\subsubsection{Step (P2)}
The presence of the indicator function $\setX\!\left(\vecz\right)$ makes it non-obvious whether this step indeed can be carried out in a decentralized fashion.
The next results show that a simple averaging procedure---analogously to that used in Step (E1) for decentralized data detection---can be carried out to perform Step (E2); the proof is given in \fref{app:P2simplification}. 

\begin{lem} \label{lem:P2simplification}
The minimization in Step (P2) simplifies to 
\begin{align} \label{eq:distributedprojection}
\bmz^{(t+1)}_c & = \textstyle \bmw^{(t)}_c + \max\!\left\{0,1-\frac{\varepsilon}{\|\bms-\vecv^{(t)}\|_2}\right\}\! \left( \frac{1}{C}\bms-\vecv^{(t)}\right)
\end{align}
with $\vecv^{(t)}=\frac{1}{C}\bmw^{(t)}=\frac{1}{C}\sum_{c=1}^C\bmw_{c}^{(t)}; \bmw_{c}^{(t)}=\bH_{c}\bmx_{c}^{(t+1)}-\bmlambda_{c}^{(t)}$.
\end{lem}

For $\varepsilon=0$, we get an even more compact expression 
\begin{align*}
 \bmz^{(t+1)}_c & = \bmw^{(t)}_c + \textstyle  \frac{1}{C} \bms-\vecv^{(t)}, \,\,c=1,\ldots,C
\end{align*}
Evidently \eqref{eq:distributedprojection} only requires a simple averaging procedure, which can be carried out by gathering local computation results from and broadcasting the averaged consensus back to each cluster. 

\subsubsection{Step (P3)}
This step can be performed independently in each cluster after distributing $\bmw^{(t)}$ and getting local $\bmz^{(t+1)}_c$. 

The resulting ADMM-based decentralized beamforming procedure is summarized in Algorithm~\ref{alg:admm_prec}, where we assume $\varepsilon=0$. To facilitate implementation of the decentralized beamforming algorithm, we initialize $\mathbf{z}_c^{(1)}, \boldsymbol\lambda_c^{(1)}, \mathbf{x}_c^{(1)}$ and then update the variables in the order of $\mathbf{z}^{(t)}_c, \boldsymbol\lambda^{(t)}_c, \mathbf{x}^{(t)}_c$ realizing that the final output of the local beamformer is simply $\mathbf{x}^{(T_\text{max})}_c$. Note that  Algorithm~\ref{alg:admm_prec} slightly deviates from the step-by-step beamforming procedure in \fref{sec:admm_prec}. {Specifically, we carry out Steps (P2), (P3), and (P1), as the beamforming output results from Step (P1). This re-ordering has no effect on the algorithm's convergence or beamforming result.} We will analyze the algorithm complexity\footnote{The matrices $\mathbf{P}_c=\mathbf{A}_c^{-1}\mathbf{H}_c^H$ or $\mathbf{P}_c=\mathbf{H}_c^H\mathbf{B}_c^{-1}$ could be precomputed to avoid recurrent computations within the ADMM iterations (at line~18 or~20 in Algorithm~\ref{alg:admm_prec}). Instead, we directly compute $\mathbf{A}_c^{-1}\mathbf{H}_c^H(\mathbf{z}_c^{(t)}+\mathbf{\lambda}_c^{(t)})$ or $\mathbf{H}_c^H\mathbf{B}_c^{-1}(\mathbf{z}_c^{(t)}+\mathbf{\lambda}_c^{(t)})$, which only requires two matrix-vector multiplications; precomputing $\mathbf{P}_c$ requires costly matrix-matrix multiplications. Hence, our complexity analysis  in \fref{sec:GPUcluster} refers to Algorithm~\ref{alg:admm_prec}.} in \fref{sec:compl} and show the reference implementation of Algorithm~\ref{alg:admm_prec} in \fref{sec:GPUcluster}.

\begin{rem}
Although we propose a decentralized scheme using CG for uplink data detection in \fref{sec:dcg_det}, a similar decentralization method of CG is not applicable in the downlink. Since we partition the uplink channel matrix $\bH$ \emph{row-wise} into~$C$ blocks, we should similarly partition the downlink channel matrix \emph{column-wise} into blocks due to the channel reciprocity; this prevents an expansion analogously to \eqref{eq:CGdecentralization}. Consequently, we focus exclusively on ADMM-based beamforming.
\end{rem}


\section{Results}
\label{sec:results}

\begin{table*}
\renewcommand{\arraystretch}{1.05}
\centering
\small
\caption{Computational Complexity of Centralized and Decentralized Data Detection and Beamforming.}
\label{tb:cplx}
\begin{tabular}{@{}lccccc@{}}
\toprule 
Algorithm & \multicolumn{2}{c}{Mode} & Preprocessing & 1st iteration & Subsequent iterations (each)\tabularnewline
\midrule 
\multirow{4}{*}{ADMM-DL} 
 & \multirow{2}{*}{$S\times S$} & TM & $2US^{2}+\frac{10}{3}S^{3}-\frac{1}{3}S$ & $4SU+4S^{2}$ & $8SU+4S^{2}+6U+1$\tabularnewline
 &  & AR & $C(2US^{2}+\frac{10}{3}S^{3}-\frac{1}{3}S)$ & $C(4SU+4S^{2})$ & $C(8SU+4S^{2}+2U)+4U+1$\tabularnewline
 & \multirow{2}{*}{$U\times U$} & TM & $2SU^{2}+\frac{10}{3}U^{3}-\frac{1}{3}U$ & $4SU+4U^{2}$ & $8SU+4U^{2}+6U+1$\tabularnewline
 &  & AR & $C(2SU^{2}+\frac{10}{3}U^{3}-\frac{1}{3}U)$ & $C(4SU+4U^{2})$ & $C(8SU+4U^{2}+2U)+4U+1$\tabularnewline
\midrule 
\multirow{4}{*}{ADMM-UL} 
 & \multirow{2}{*}{$S\times S$} & TM & $2US^{2}+\frac{10}{3}S^{3}+4US+4S^{2}-\frac{1}{3}S$ & $2U$ & $8SU+4S^{2}+4U$\tabularnewline
 &  & AR & $C(2US^{2}+\frac{10}{3}S^{3}+4US+4S^{2}-\frac{1}{3}S)$ & $2U$ & $C(8SU+4S^{2}+2U)+2U$\tabularnewline
 & \multirow{2}{*}{$U\times U$} & TM & $2SU^{2}+\frac{10}{3}U^{3}+4SU+4U^{2}-\frac{1}{3}U$ & $2U$ & $4U^{2}+6U$\tabularnewline
 &  & AR & $C(2SU^{2}+\frac{10}{3}U^{3}+4SU+4U^{2}-\frac{1}{3}U)$ & $2U$ & $C(4U^{2}+4U)+2U$\tabularnewline
\midrule 
\multirow{2}{*}{CG-UL} 
 & \multirow{2}{*}{ } & TM & $4SU+2U$ & $8SU+6U$ & $8SU+12U$\tabularnewline
 &  & AR & $4CSU+2U$ & $C(8SU+4U)+2U$ & $C(8SU+10U)+2U$\tabularnewline
\midrule 
\multirow{2}{*}{Centralized} & \multicolumn{2}{c}{ZF-DL} & \multicolumn{3}{c}{$6CSU^{2}+\frac{10}{3}U^{3}+4CSU-\frac{4}{3}U$}\tabularnewline
 & \multicolumn{2}{c}{MMSE-UL} & \multicolumn{3}{c}{$6CSU^{2}+\frac{10}{3}U^{3}+4CSU-\frac{1}{3}U$}\tabularnewline
\bottomrule 
\end{tabular}
\vspace{-3mm}
\end{table*}

\begin{figure*}[t]
\subfigure[{Error-rate performance of decentralized ADMM uplink data detection.}]{\includegraphics[width=0.999\textwidth]{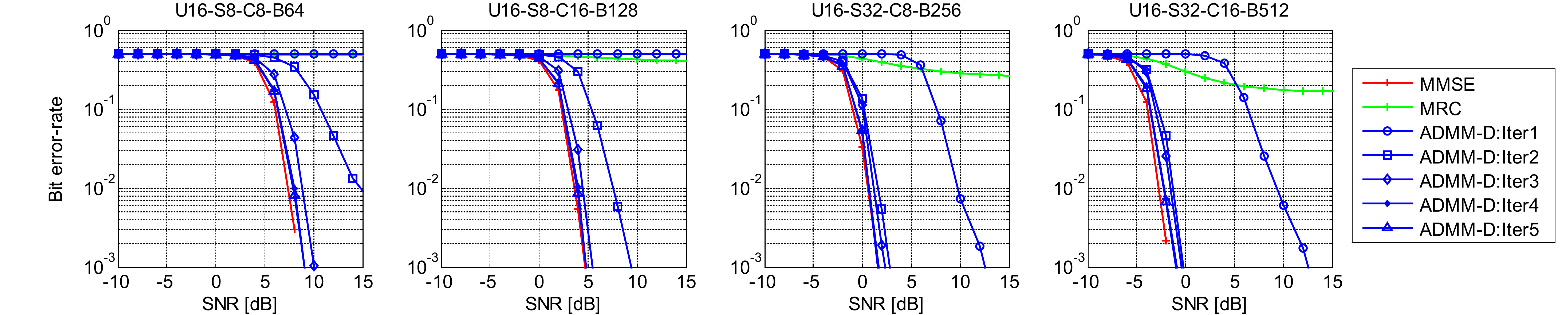}\label{fig:BER_admm_d}}
\subfigure[Error-rate performance of decentralized CG uplink data detection.]{\includegraphics[width=0.999\textwidth]{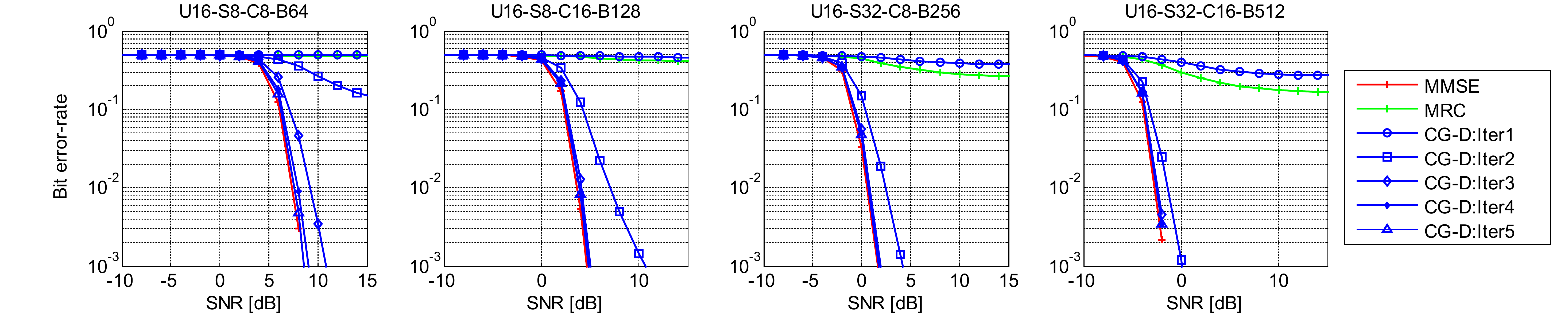}\label{fig:BER_cg_d}}
\subfigure[Error-rate performance of decentralized ADMM downlink beamforming.]{\includegraphics[width=0.999\textwidth]{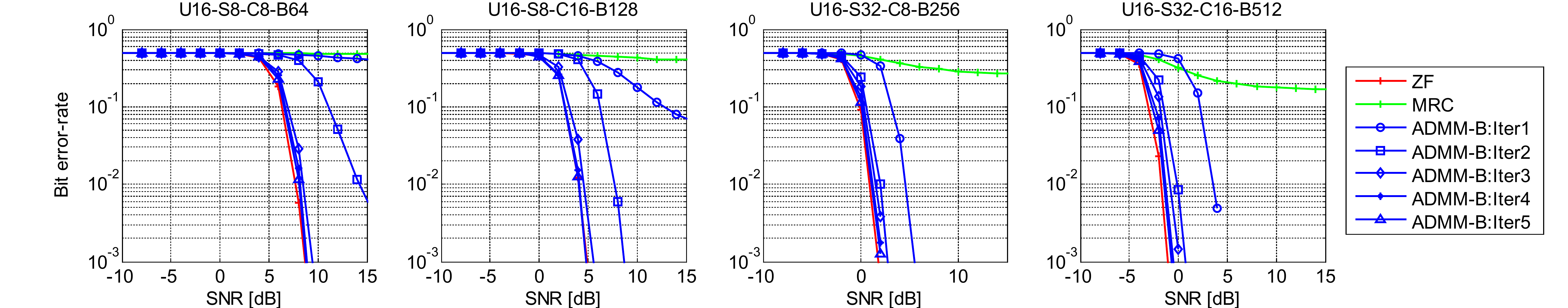}
\label{fig:BER_admm_b}}
%
\caption{\revision{Bit error-rate (BER) performance of decentralized data detection and beamforming; we use the notation $U-S-C-B$ (representing the number of users $U$, antennas per clusters $S$,  clusters $C$, and BS antennas $B$) as subtitle of each figure to indicate the corresponding system configuration.}}
\label{fig:BERfig}
\vspace{-2mm}
\end{figure*}

We now analyze the computational complexity and consensus bandwidth of our proposed algorithms. We also show error-rate simulation results in LTE-like massive MU-MIMO uplink and downlink systems. We investigate the performance/complexity trade-offs and show  practical operating points of our decentralized methods under various antenna configurations, providing design guidelines for decentralized massive MU-MIMO BSs.

\subsection{Computational Complexity}
\label{sec:compl}

In Table \ref{tb:cplx}, we list the number of real-valued multiplications\footnote{We ignore data-dependencies or other operations, such as additions, divisions, etc. While this complexity measure is rather crude, it enables insights into the pros and cons of decentralized baseband processing.} of our decentralized ADMM-based downlink beamforming (ADMM-DL), ADMM-based uplink detection (ADMM-UL) and CG-based uplink detection (CG-UL) algorithms.
We also compare the complexity to that of conventional, centralized ZF downlink beamforming (ZF-DL) and MMSE uplink detection (MMSE-UL). For all decentralized algorithms and modes, for example, the ``$S\times S$ mode'' when $S\leq U$ and the ``$U\times U$ mode'' when $S>U$, we show the \emph{timing (TM) complexity} and \emph{arithmetic (AR) complexity}. 
We assume that the centralized computations take place on a centralized PE while decentralized computations are carried out on multiple decentralized PEs. For the centralized computations, both the TM and AR complexities count the number of real-valued multiplications on the centralized PE. For the decentralized operations, the TM complexity only counts operations that take place on \emph{a single} local processing unit where all decentralized local processors perform their own computations \emph{in parallel} at the same time, thus reflecting the latency of algorithm execution; in contrast, the AR complexity counts the \emph{total} complexity accumulated from \emph{all} local processing units, thus reflecting the total hardware costs.
The complexity of our methods depends on the number of clusters $C$, the number of users $U$, the number of BS antennas $S$ per antenna cluster, and the number of iterations $T_\text{max}$ to achieve satisfactory error-rate performance. We also divide the complexity counts into three parts: preprocessing before ADMM or CG iterations, first iteration, and subsequent iterations. The complexity in the first iteration is typically lower as many vectors are zero. 

Table \ref{tb:cplx} reveals that preprocessing for ADMM exhibits relatively high complexity, whereas CG-based detection is computationally efficient. The per-iteration complexity of ADMM is, however, extremely efficient (depending on the operation mode). Overall, CG-based data detection is more efficient than the ADMM-based counterpart, whereas the latter enables more powerful regularizers. 
Centralized ZF or MMSE beamforming or detection, respectively, require high complexity, i.e., scaling with $U^3$, but generally achieve excellent error-rate performance~\cite{mimo_next_gen}. We will analyze the trade-offs between complexity and error-rate performance in \fref{sec:tradeoff_sim}.

\subsection{Consensus Bandwidth}
The amount of data passed between the centralized processing unit and the decentralized local units during ADMM or CG iterations scales with the dimension of the consensus vector $\bmw^{(t)}$. For a single subcarrier, $\bmw^{(t)}$ is a vector with $U$ complex-valued entries. If we perform detection or precoding for a total of $N_\text{CR}$ subcarriers, then in each ADMM or CG iteration, we need to gather $U\times N_\text{CR}$ complex-valued entries from each local processing unit for consensus vectors corresponding to $N_\text{CR}$ subcarriers, and broadcast all $N_\text{CR}$ consensus vectors to each local processor afterwards. Such a small amount of data exchange relaxes the requirement on interconnection bandwidth among decentralized PEs, and avoids the large data transfer between the entire BS antenna array and BS processor in a conventional centralized BS. However, as we will show with our GPU cluster implementation in \fref{sec:GPUcluster}, the interconnect \emph{latency} of the network critically effects the throughput of DBP. 

\subsection{Error-Rate Performance}
\label{sec:ber_sim}
We simulate our decentralized data detection and beamforming algorithms in an LTE-based large-scale MIMO  system.\footnote{\revision{A simplified MATLAB simulator for DBP in the uplink and downlink is available on GitHub at \url{https://github.com/VIP-Group/DBP}}} For both the uplink and downlink simulations, we consider OFDM transmission with 2048 subcarriers in a  20\,MHz channel, and incorporate our algorithms with other necessary baseband processing blocks, including 16-QAM modulation with Gray mapping, FFT/IFFT for subcarrier mapping, rate-5/6 convolutional encoding with random interleaving and soft-input Viterbi-based channel decoding~\cite{SFBQ12}. We generate the channel matrices using the Winner-II channel model~\cite{winner} and consider channel estimation errors, i.e., we assume a single orthogonal training sequence per user and active subcarrier. {For the sake of simplicity, we avoid rate adaptation, the use of cyclic redundancy checks, and (hybrid) ARQ transmission protocols.}

In \fref{fig:BERfig}, we show the coded bit error-rate (BER) performance against average SNR per receive antenna for decentralized ADMM detection (\fref{fig:BER_admm_d}), for decentralized CG detection (\fref{fig:BER_cg_d}) in the uplink, and for decentralized ADMM beamforming (\fref{fig:BER_admm_b}) in the downlink. We consider various antenna configurations. We fix the number of users $U=16$, and set $S=8$ (for $S\leq U$ case) or $S=32$ (for $S>U$ case), and scale the total BS antenna number $B=S\times C$ from 64 to 512 by choosing $C=8$ and $C=16$. 

We see that for all the considered antenna and cluster configurations, only 2-to-3 ADMM or CG iterations are sufficient to approach the performance of the linear MMSE equalizer. For the $S>U$ case, even a single ADMM iteration enables excellent BER performance for detection and beamforming without exhibiting an error floor{, which outperforms CG with one iteration}. 
We note that the amount of consensus information that must be exchanged during each ADMM or CG iteration is rather small. {Hence, our decentralized data detection and beamforming algorithms are able to achieve the error-rate performance of centralized solutions (such as MMSE and MRC data detection or ZF beamforming) without resulting in prohibitive interconnect or I/O bandwidth}---this approach enables highly scalable and modular BS designs with hundreds or thousands of antenna elements. 

\begin{figure*}[t]

\subfigure[{UL ADMM-based data detection.}]{\includegraphics[width=0.33\textwidth]{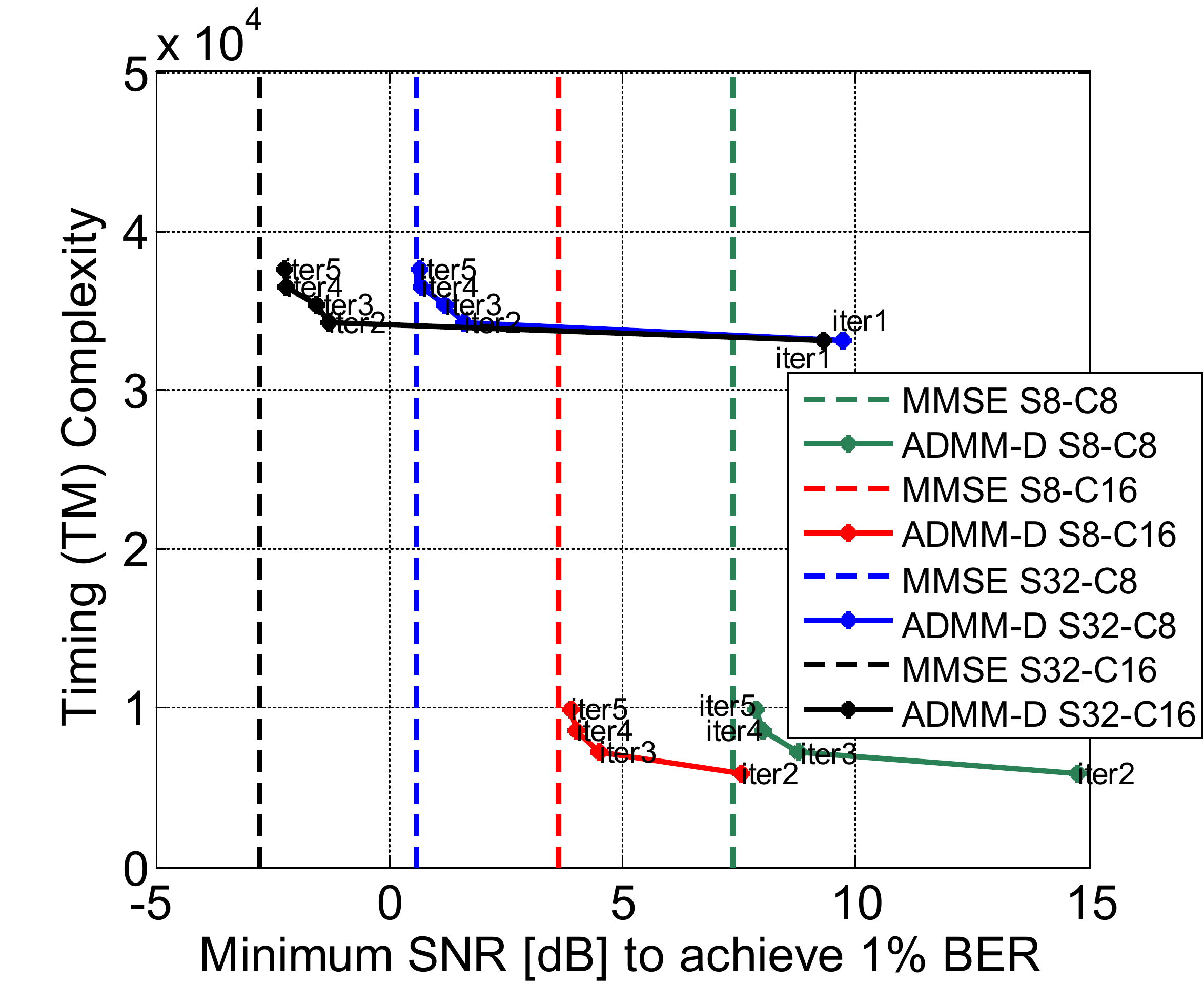}\label{fig:TD_admm_d}}\hspace{0.1cm}
\subfigure[UL decentralized CG-based data detection.]{\includegraphics[width=0.33\textwidth]{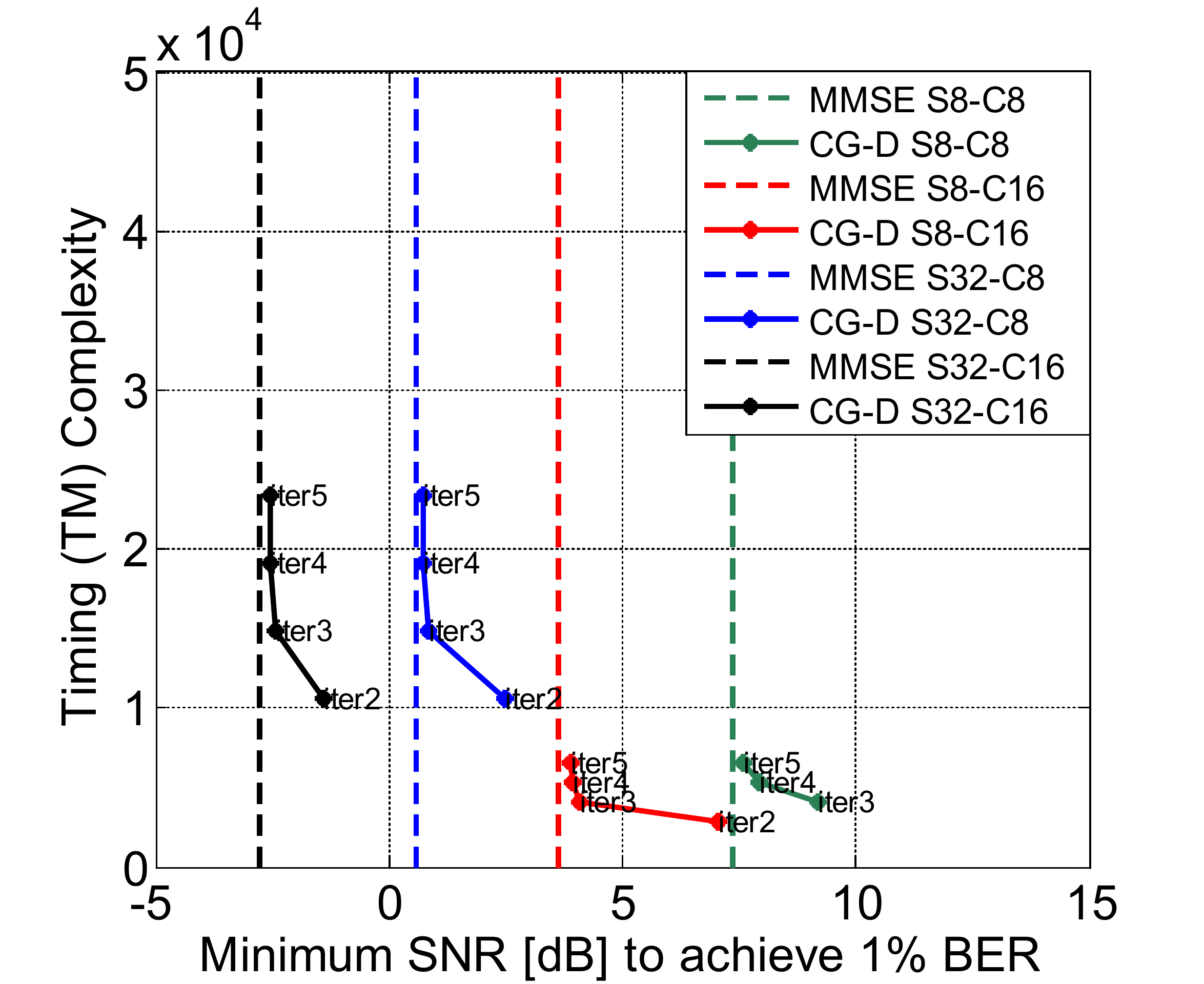}\label{fig:TD_cg_d}}\hspace{0.1cm}
\subfigure[DL ADMM-based beamforming.]{\includegraphics[width=0.33\textwidth]{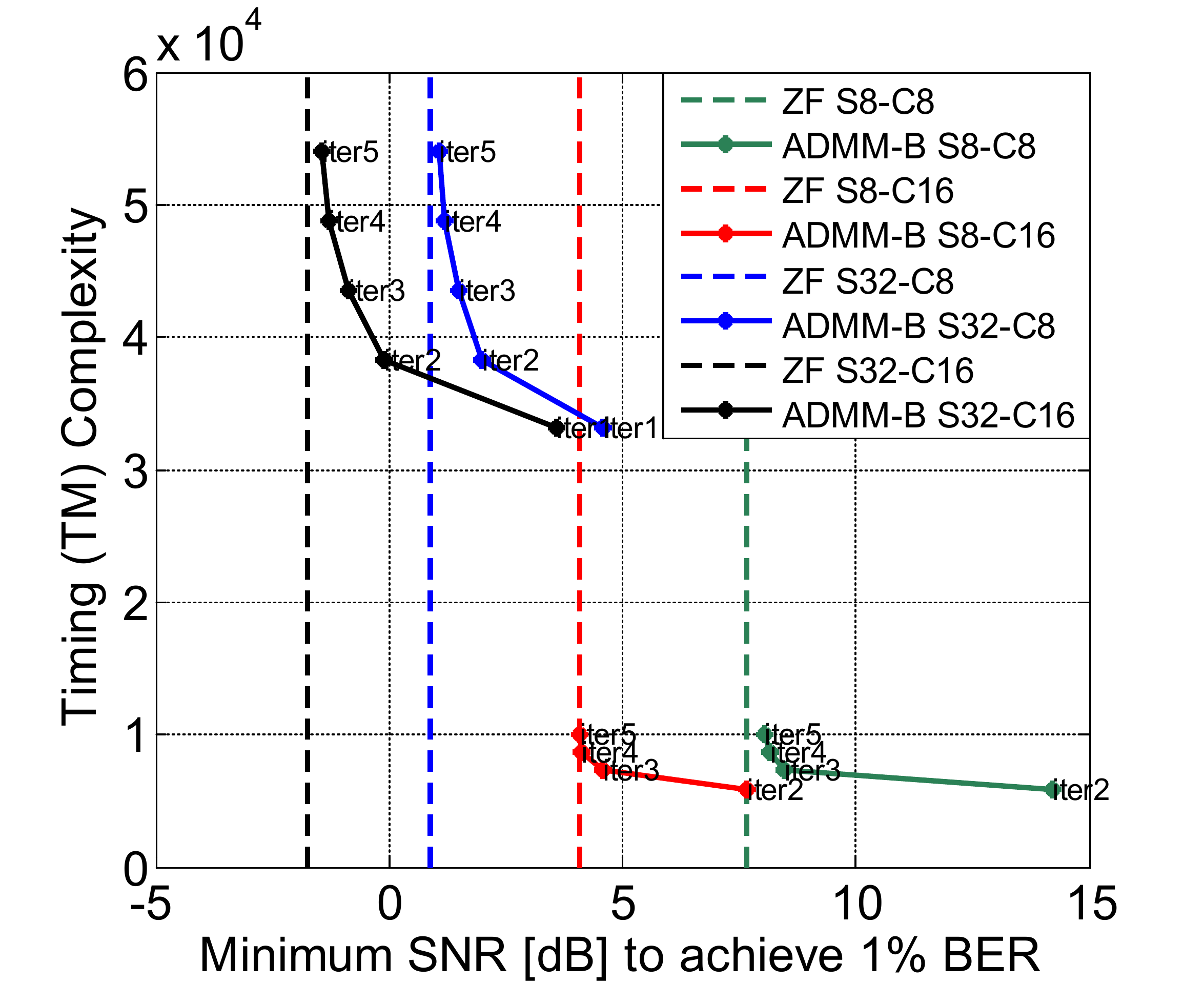}\label{fig:TD_admm_b}}
\vspace{-0.2cm}
\caption{Performance/complexity trade-off of decentralized data detection and beamforming in an LTE-like massive MU-MIMO system with $U=16$ users.}
\label{fig:tradeoffplot}
\end{figure*}

\subsection{Performance/Complexity Trade-off Analysis}
\label{sec:tradeoff_sim}
\fref{fig:tradeoffplot} illustrates the trade-off between error-rate performance and computational complexity of our proposed methods. 
As a performance metric, we consider the minimum required SNR to achieve $1\%$ BER; the complexity is characterized by the TM complexity and depends on the number of ADMM or CG iterations (the numbers next to the curves). As a reference, we also include the BER performance of centralized MMSE data detection and ZF beamforming (dashed vertical lines).

For the uplink, Figures~\ref{fig:TD_admm_d} and \ref{fig:TD_cg_d} show the trade-offs for ADMM-based and CG-based data detection, respectively. We see that only a few CG iterations are sufficient to achieve near-MMSE performance whereas ADMM requires a higher number of iterations to achieve the same performance. CG-based data detection exhibits the better trade-off here, and is the preferred method. However, for scenarios such as $U<S$, ADMM-based detection exhibits no error floor, even for a single iteration, while CG-based data detection performs rather poorly at one iteration. In addition, our ADMM-based method supports more sophisticated regularizers (such as the BOX regularizer).  

For the downlink, \fref{fig:TD_admm_b} shows our proposed ADMM-based beamformer. We see that only a few iterations (e.g., 2 to 3 iterations) are necessary to achieve near-optimal performance.  In addition, for small antenna cluster sizes (e.g., $S=8$), the complexity is comparable to CG-based detection; for large antenna cluster sizes, the complexity is only $2\times$ higher.


\section{GPU Cluster Implementation}
\label{sec:GPUcluster}
\begin{figure}[tp]
\centering
\includegraphics[width=0.99\columnwidth]{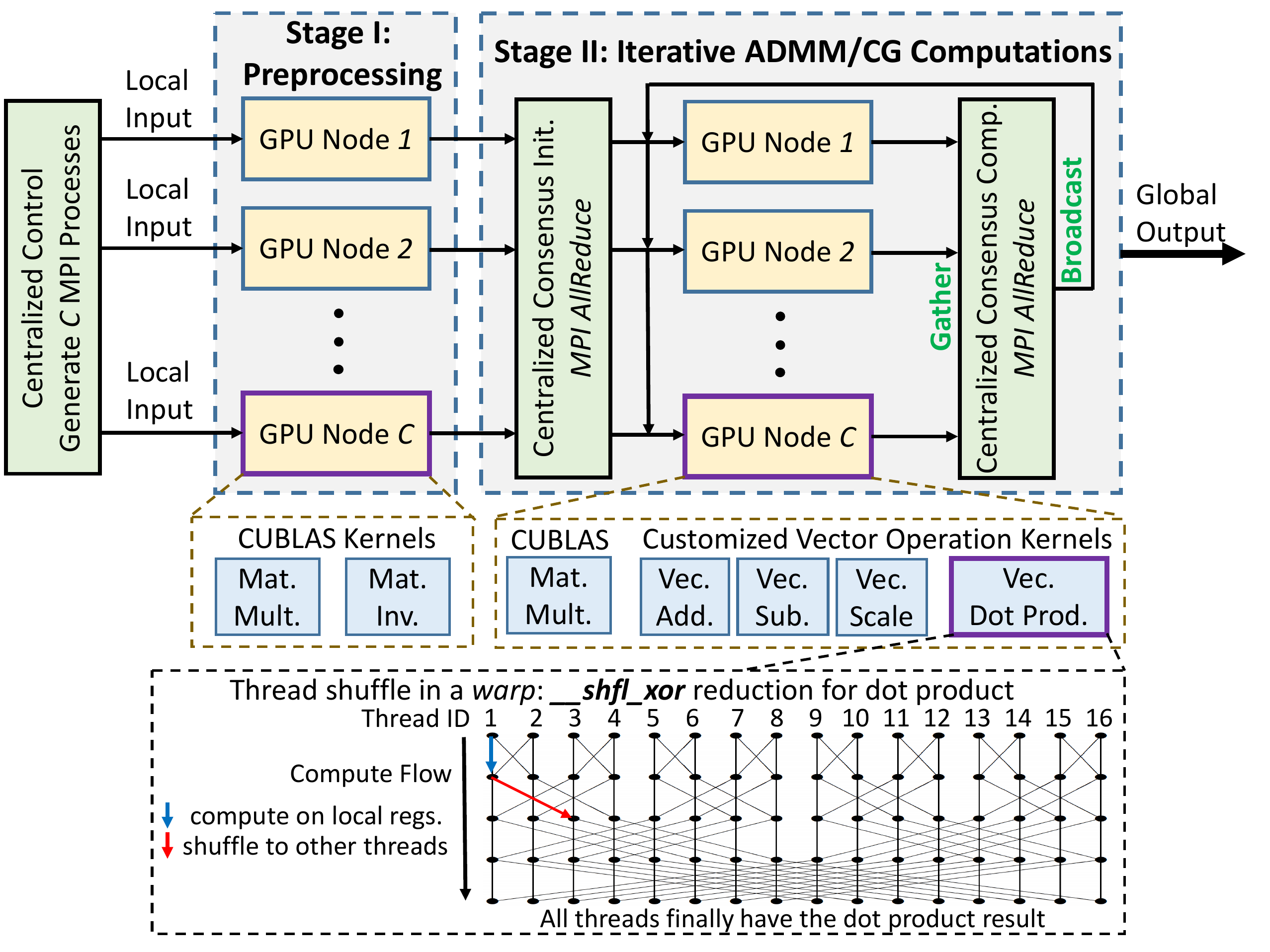}
\vspace{-0.5cm}
\caption{{Mapping of the algorithms on the GPU cluster. Each GPU node performs local data detection (uplink) or beamforming (downlink) on $C$ CUDA GPUs using CUBLAS or customized kernel blocks as detailed in Algorithms 1, 2, and 3. The bottom part illustrates the dot product kernel, in which we use a \emph{warp shuffle}\cite{cuda} approach to minimize the processing latency.}}
\label{fig:gpu_map}
\end{figure}

General purpose computing on GPU (GPGPU) is widely used for fast prototyping of baseband algorithms in the context of reconfigurable software-defined radio (SDR) systems~\cite{cg_gpu,sdr_gpu,mimodet_gpu}. \revision{We now present  reference implementation results of the proposed decentralized data detection and beamforming algorithms on a GPU cluster to demonstrate the practical scalability of DBP in terms of throughput.} We consider a wideband scenario, which enables us to exploit decentralization across subcarriers and in the BS antenna domain. 
{Fig. \ref{fig:gpu_map} illustrates the mapping of our algorithms onto the GPU cluster, the main data flow, and the key computing modules.} 
For all our implementations, we use the message passing interface (MPI) library~\cite{mpi} to generate~$C$ independent processes on~$C$ computing nodes in the GPU cluster, where each process controls a GPU node for accelerating local data detection or beamforming using CUDA~\cite{cuda}. 
{Data collection and broadcasting among GPUs nodes can be realized by MPI function calls over a high-bandwidth Cray Aries~\cite{aries} or Infiniband~\cite{infiniband} interconnect network.}
\revision{We benchmark our implementations for a variety of antenna and cluster configurations to showcase the efficacy and scalability of DBP to very large BS antenna arrays with decentralized computing platforms.}

\subsection{Design Mapping and Optimization Strategies}
We next discuss the implementation details and optimizations that achieve high throughput with our decentralized algorithms.

\subsubsection{Optimizing kernel computation performance}
The local data detection and beamforming computations in each cluster are mapped as GPU kernel functions, which can be invoked with thousands of threads on each GPU node to realize inherent algorithm parallelism and to exploit the massive amount of computing cores and memory resources. 
All of our decentralized algorithms mainly require matrix-matrix and matrix-vector multiplications. The ADMM methods also involve an explicit matrix inversion step. Such computations are performed efficiently using the \emph{cuBLAS} library~\cite{cublas}, a CUDA-accelerated basic linear algebra subprograms (BLAS) library for GPUs.
We use the \texttt{cublasCgemmBatched} function to perform matrix-matrix multiplications and matrix-vector multiplications, and use \texttt{cublasCgetrfBatched} and \texttt{cublasCgetriBatched} to perform fast matrix inversions via the Cholesky factorization followed by forward-backward substitution~\cite{matcomp}. 
For these functions, ``\texttt{C}'' implies that we use complex-valued floating point numbers and ``\texttt{Batched}'' indicates that the function can complete a batch of computations in parallel, which are scaled by the \texttt{batchsize} parameter of the function with a single function call. 
Since the local data detection or beamforming problems are solved independently for each subcarrier, we can group a batch of subcarriers and process them together to achieve high GPU utilization and throughput. For each data detection or  beamforming computation cycle, we define $N_\text{sym}$ OFDM symbols, each including $N_\text{sc}$ subcarriers, as the total workload to be processed by such \texttt{Batched} kernel function calls. 
We assume that the channel remains static for~$N_\text{sym}$ symbols. 

For the preprocessing stage, the matrix-matrix multiplications and matrix inversions, which only depend on $\mathbf{H}_c$, can be calculated with $\texttt{batchsize} \!=\!N_\text{sc}$ for $N_\text{sc}$ subcarriers in an OFDM symbol, and then broadcast to all $N_\text{sym}$ symbols inside GPU device memory to reduce complexity. {For the matrix-vector multiplications, we invoke \emph{cuBLAS} functions with $\texttt{batchsize}\!=\!N_\text{sc}\times N_\text{sym}$, because these computations depend on transmit or receive symbols as well.}

For all other types of computations, such as vector addition/subtraction and inner product calculations,  we use customized kernel functions. In this way, we can combine several vector computation steps into a single kernel, and take advantage of local registers or shared memories to store and share intermediate results instead of using slower GPU device memories and multiple \emph{cuBLAS} functions. 
Vector addition/subtraction for $N_\text{sc}\times N_\text{sym}$ number of $U$-element vectors exposes explicit data parallelism and can proceed with $N_\text{sc}\times N_\text{sym}\times U$ GPU threads in parallel. However, the dot product for each pair of $U$-dimensional vectors requires internal communication among a group of $U$ threads, each thread controlling an element-wise multiplication, to be reduced to a sum. 
A typical way for such a reduction is to resort to the shared memory, an on-chip manageable cache with L1 cache speed, where a certain group of $U$ threads associated with a vector sums up their element-wise multiplication results to a shared variable atomically, for example, using the \texttt{atomicAdd} CUDA function call. However, shared memory typically suffers from higher latency (compared to that of local registers) and also from possible resource competition among multiple threads. In the CG-based data detector, we utilize the \emph{warp shuffle} technique, which is supported by Kepler-generation GPUs for efficient register-to-register data shuffling among threads within a thread \emph{warp}\cite{cuda}, for faster parallel reduction. {As shown in Fig. \ref{fig:gpu_map}, we use the \texttt{\_\_shfl\_xor(var,laneMask)} intrinsic,  which retrieves the register value of variable \texttt{var} from a certain thread with lane ID \emph{source\_id} for the calling thread with lane ID $dest\_id$ within the same \emph{warp}, where $source\_id$ satisfies: $source\_id$ XOR $dest\_id=\texttt{laneMask}$, XOR indicating bitwise exclusive or. In this way, the parallel sum reduction for the inner product of two $U$-element vectors can be realized by \texttt{var=\_\_shfl\_xor(val,laneMask)+var} in $\log_2(U)$ iterations on a reduction tree with initial $\texttt{laneMask}=U$, and \texttt{laneMask} reduced to half in each iteration. Finally, as shown in Fig. \ref{fig:gpu_map}, each of the $U$ threads will have a copy of the inner-product result stored in its own \texttt{var}, i.e., the above process is actually an operation of \emph{allreduce} rather than \emph{reduce}, which facilitates downstream computations in which each thread requires the value of the inner product. Here, we assume that the number of user antennas satisfies $U\leq\textit{warpsize}=32$ and is a power of two, for example, $U=8$ or $U=16$. Otherwise, we can resort to the alternative vector inner product solution by  \texttt{atomicAdd} using shared memory.}
The optimizations described above enable the computation of each iteration using $N_\text{sc}\times N_\text{sym}\times U$ threads using fast on-chip memory resources and efficient inter-thread communication schemes that avoid blocking.

\begin{table}
\renewcommand{\arraystretch}{1.05}
\small
\caption{{Parallelism Analysis and Mapping Strategies.}}
\label{tb:parallelism}
\begin{center}
\begin{tabular}{@{}lccc@{}}
\toprule 
{Module} & {Strategy} & {Parallelism} & {Memory$^{\footnotesize1}$}\tabularnewline
\midrule 
{Mat.\ mult.} & {batch cuBLAS} & {$N_\text{sc}\times N_\text{sym}$}  & {g,c,s,r} \tabularnewline
{Mat.\ inv.} & {batch cuBLAS} & {$N_\text{sc}\times N_\text{sym} $} & {g,c,s,r} \tabularnewline
{Vec.\ +/-/scale} & {multi-threading} & {$N_\text{sc}\times N_\text{sym} \times U$} & {g,c,r}\tabularnewline
{Vec.\ dot prod.} & {warp shuffle} & {$N_\text{sc}\times N_\text{sym}\times U$} & {g,c,r}\tabularnewline
{GPU comm.} & {MPI, RDMA} & {Among $C$ GPUs} & {g} 
\tabularnewline
\bottomrule
\end{tabular}
\end{center}
\begin{tablenotes}
\item[1] {$^{\footnotesize1}$g: global (device) memory; c: cache; s: shared memory, r: register}
\end{tablenotes}
\end{table}

\subsubsection{Improving message passing efficiency}
Message passing latency is critical to the efficiency of our design. For our decentralized algorithms, the consensus operations require data collection and sharing among $C$ GPU nodes. This can be realized by MPI \emph{collective} function calls for inter-process communication among $C$ controlling processes with messages of size $N_\text{sc}\times N_\text{sym}\times U$ complex samples. 
More specifically, we choose the  \texttt{MPI\_Allreduce} function to sum (and then average) vectors across nodes. We then broadcast the resulting consensus vector $\bmw$ to all local nodes within this single collective MPI function call. 
Typically, \texttt{MPI\_Allreduce} operates on the CPU's memory and requires GPU arrays to be copied into a CPU memory buffer before calling \texttt{MPI\_Allreduce}. To eliminate redundant memory copy operations, we take advantage of \emph{CUDA-aware MPI}~\cite{cudaaware} and \emph{GPUDirect} remote device memory access (RDMA) techniques~\cite{gpudirect}, which enable the MPI function to explicitly operate on GPU memories without using a CPU intermediary buffer.  This results in reduced latency and higher bandwidth.
{Table \ref{tb:parallelism} summarizes the key mapping strategies, degrees of parallelism, and associated memory usage for both intra-GPU computing modules and inter-GPU communication mechanisms of our GPU implementations.}

\begin{table*}[tp]
\renewcommand{\arraystretch}{1.05}
\small
\centering
\caption{Latency (L) in [ms] and Throughput (T) in [Mb/s] for Decentralized Data Detection and Beamforming ($U=16$).}
\setlength\tabcolsep{3pt}
\label{tb:implementationresults}
\scalebox{0.97}{ 
\begin{tabular}{@{}lccccccccc@{}}
\toprule 
\revision{$B$}  & \revision{64} &  \revision{128} & \revision{256} & \revision{128} & \revision{256} & \revision{512} & \revision{256} & \revision{512} & \revision{1024} \tabularnewline
\revision{$C$} & \revision{8} & \revision{16} & \revision{32} & \revision{8} & \revision{16} & \revision{32} & \revision{8} & \revision{16} & \revision{32} \tabularnewline
\revision{$S$} & \revision{8} & \revision{8} & \revision{8} & \revision{16} & \revision{16} & \revision{16} & \revision{32} & \revision{32} & \revision{32} \tabularnewline
\midrule 
\revision{Iter.} & \revision{L / T} & \revision{L / T} & \revision{L / T} & \revision{L / T} & \revision{L / T} & \revision{L / T} & \revision{L / T} & \revision{L / T} & \revision{L / T}\tabularnewline
\midrule 
\multicolumn{10}{c}{\revision{ADMM-based decentralized uplink data detection}} \\
\midrule 
\revision{1} & \revision{2.060 / 417.5} & \revision{2.166 / 397.1} & \revision{2.520 / 341.4}  & \revision{3.810 / 225.7} & \revision{4.079 / 210.9} & \revision{4.466 / 192.6} & \revision{4.329 / 198.7} & \revision{4.516 / 190.5} & \revision{4.693 / 183.3}\tabularnewline
\revision{2} & \revision{4.989 / 172.4} & \revision{5.411 / 159.0} & \revision{6.451 / 133.3} & \revision{6.756 / 127.3} & \revision{7.597 / 113.2} & \revision{8.495 / 101.3} & \revision{7.173 / 119.9} & \revision{7.855 / 109.5} & \revision{8.615 / 99.84}\tabularnewline
\revision{3} & \revision{7.728 / 111.3} & \revision{8.561 / 100.5} & \revision{9.910 / 86.80} & \revision{9.712 / 88.56} & \revision{10.94 / 78.62} & \revision{12.68 / 67.85} & \revision{10.22 / 84.15} & \revision{11.27 / 76.30} & \revision{12.98 / 66.28}\tabularnewline
\revision{4} & \revision{10.80 / 79.67} & \revision{11.92 / 72.16}  & \revision{13.79 / 62.39} & \revision{12.44 / 69.12} & \revision{14.08 / 61.10} & \revision{16.64 / 51.68} & \revision{13.49 / 63.78} & \revision{14.83 / 57.99} & \revision{17.27 / 49.81}\tabularnewline
\revision{5} & \revision{13.43 / 64.01} & \revision{15.69 / 54.83} & \revision{17.59 / 48.90} & \revision{15.18 / 56.66} & \revision{17.88 / 48.10} & \revision{21.11 / 40.75} & \revision{16.50 / 52.13} & \revision{18.65 / 46.12} & \revision{21.53 / 39.95}\tabularnewline
\midrule
\multicolumn{10}{c}{\revision{CG-based decentralized uplink data detection}} \\
\midrule 
\revision{1} & \revision{3.516 / 244.7} & \revision{4.077 / 211.0} & \revision{4.594 / 187.2} & \revision{3.811 / 225.7} & \revision{4.325 / 198.9} & \revision{4.960 / 173.4} & \revision{4.232 / 203.3} & \revision{4.729 / 181.9} & \revision{4.984 / 161.8}\tabularnewline
\revision{2} & \revision{5.078 / 169.4} & \revision{6.192 / 138.9} & \revision{6.574 / 130.8} & \revision{5.597 / 153.7} & \revision{6.160 / 139.6} & \revision{7.190 / 119.6} & \revision{6.207 / 138.6} & \revision{7.067 / 121.7} & \revision{7.150 / 112.8}\tabularnewline
\revision{3} & \revision{6.567 / 131.0} & \revision{8.250 / 104.3} & \revision{8.490 / 101.3} & \revision{7.323 / 117.5} & \revision{8.436 / 102.0} & \revision{9.310 / 92.39} & \revision{7.841 / 109.7} & \revision{9.100 / 94.52} & \revision{9.314 / 86.58}\tabularnewline
\revision{4} & \revision{8.080 / 106.5} & \revision{10.02 / 85.87} & \revision{10.79 / 79.70} & \revision{9.243 / 93.06} & \revision{10.22 / 84.12} & \revision{11.91 / 72.20} & \revision{9.775 / 88.00} & \revision{10.90 / 78.94} & \revision{11.45 / 70.43}\tabularnewline
\revision{5} & \revision{9.787 / 87.88} & \revision{11.97 / 71.85} & \revision{12.74 / 67.52} & \revision{11.07 / 77.69} & \revision{11.95 / 71.99} & \revision{14.03 / 61.30} & \revision{11.93 / 72.08} & \revision{13.27 / 64.82} & \revision{13.61 / 59.25}\tabularnewline
\midrule 
\multicolumn{10}{c}{\revision{ADMM-based decentralized downlink beamforming}} \tabularnewline
\midrule 
\revision{1} & \revision{0.744 / 1156} & \revision{0.745 / 1155} & \revision{0.746 / 1154} & \revision{2.261 / 380.5} & \revision{2.266 / 379.6} & \revision{2.271 / 378.8} & \revision{2.785 / 308.8} & \revision{2.792 / 308.1} & \revision{2.797 / 307.6}\tabularnewline
\revision{2} & \revision{2.685 / 320.3} & \revision{2.849 / 301.9} & \revision{2.974 / 289.2} & \revision{4.048 / 212.5} & \revision{4.318 / 199.2} & \revision{4.557 / 188.8} & \revision{4.618 / 186.3} & \revision{4.762 / 180.6} & \revision{5.084 / 169.2}\tabularnewline
\revision{3} & \revision{4.567 / 188.4} & \revision{4.954 / 173.6} & \revision{4.803 / 179.1} & \revision{5.741 / 149.8} & \revision{5.966 / 144.2} & \revision{6.240 / 137.9} & \revision{6.323 / 136.0} & \revision{6.651 / 129.3} & \revision{7.213 / 119.2}\tabularnewline
\revision{4} & \revision{6.269 / 137.2} & \revision{6.797 / 126.6} & \revision{6.769 / 127.1} & \revision{7.352 / 117.0} & \revision{7.741 / 111.1} & \revision{8.720 / 98.65} & \revision{7.849 / 109.6} & \revision{8.371 / 102.8} & \revision{9.380 / 91.70}\tabularnewline
\revision{5} & \revision{8.055 / 106.8} & \revision{9.012 / 95.45} & \revision{8.753 / 98.27} & \revision{8.766 / 98.12} & \revision{9.558 / 89.99} & \revision{10.42 / 82.54} & \revision{9.570 / 89.88} & \revision{10.15 / 84.76} & \revision{11.11 / 77.40}\tabularnewline
\midrule 
\multicolumn{10}{c}{{\revision{Centralized MMSE uplink data detection and ZF downlink beamforming (baseline implementations)}}} \tabularnewline
\midrule 
{\revision{MMSE}} & \revision{3.755 / 229.1} & \revision{5.371 / 160.2} & \revision{8.607 / 99.94} & \revision{5.371 / 160.2} & \revision{8.607 / 99.94} & \revision{15.04 / 57.20} & \revision{8.607 / 99.94} & \revision{15.04 / 57.20} & \revision{27.97 / 30.75}\tabularnewline
{\revision{ZF}} & \revision{4.063 / 211.7} & \revision{5.936 / 144.9} & \revision{9.698 / 88.70} & \revision{5.936 / 144.9} & \revision{9.698 / 88.70} & \revision{17.18 / 50.06} & \revision{9.698 / 88.70} & \revision{17.18 / 50.06} & \revision{32.19 / 26.72}\tabularnewline
\bottomrule 
\end{tabular}
}
\end{table*}

\subsection{Implementation Results}
We implemented our algorithms on a Navy DSRC Cray XC30 cluster~\cite{navy} equipped with $32$ GPU nodes connected by a Cray Aries network interface. Each node has a $10$-core Intel Xeon E5-2670v2 CPU and an Nvidia Tesla K40 GPU with 2880 CUDA cores and 12\,GB GDDR5 memory. \revision{The Cray Aries network uses a novel Dragonfly topology that enables fast and scalable network communication with a peak all-to-all global bandwidth of 11.7\,GB/s per node for the full network~\cite{aries}. The hybrid MPI and CUDA designs are compiled by Nvidia's nvcc compiler and the Cray compiler, and linked with CUDA's runtime library, the \emph{cuBLAS} library, and the Cray MPICH2 library. Our software implementations can be easily reconfigured with new design parameters, such as number of BS antennas, modulation schemes, etc, and recompiled in a few seconds to enable high design flexibility.
In what follows, we benchmark the latency (in milliseconds) and throughput (in Mb/s) of our implementations based on CPU wall-clock time.}

Table~\ref{tb:implementationresults} summarizes the latency and throughput performance of ADMM-based decentralized data detection, CG-based decentralized data detection, and ADMM-based decentralized beamforming, depending on the number of iterations $T_\text{max}$. \revision{We also include the performance of \emph{centralized} MMSE data detection and ZF beamforming designs based on our previous results reported in \cite{cg_gpu} as a baseline.\footnote{\revision{Centralized data detectors and beamformers for massive MU-MIMO have been implemented on FPGAs and ASICs in, e.g., \cite{prabhu20173,wu2014large,yin20143}. A direct and fair comparison with our GPU implementations is, however, difficult.}}}
We consider a scenario with 64-QAM, a coherence interval of $N_\text{sym}=7$ symbols, and $N_\text{sc}=1200$ active subcarriers, which reflects a typical slot of a 20\,MHz LTE frame. 
\revision{We fix the number of users to $U=16$ and show results for three scenarios: (i) $U > S$ with $S=8$, (ii) $U = S = 16$, and (iii) $U < S$ with $S=32$. For each scenario, we vary the number of BS antennas as $B=C S$ for different cluster sizes $C\in\{8,16,32\}$. 
The measured latency includes both kernel-computation and inter-GPU message-passing latencies. The computation latency scales up with local computation workload, while the average message-passing latency is approximately $1\sim 2$\,ms in each ADMM or CG iteration  
and remains nearly constant for $C\leq32$ thanks to the scalable Dragonfly topology of Cray Aries.}

We see that by increasing the number of clusters $C,$ and hence the total number of BS antennas $B$, the achieved throughput degrades only slightly; this demonstrates the excellent scalability of DBP to large antenna arrays. In stark contrast, centralized MMSE and ZF methods suffer from an orders-of-magnitude throughput degradation when increasing the number of BS antennas; this clearly shows the limits of centralized data detection and beamforming methods. \revision{We also note that for MIMO systems with a relatively small number of BS antennas, such as, when $B=64$ or $B=128$, we see that centralized data detection and precoding is able to achieve a higher throughput than decentralized schemes. We emphasize, however, that centralized processing assumes that one is able to get the raw baseband data into the single, centralized computing fabric at sufficiently high data rates.}
We furthermore we see that for a given number of clusters $C$, the throughput for the $S=32$ case is smaller than that of the $S=8$ case. The reason for this behavior is the fact that having a large number of antennas per cluster $S$ leads to a higher complexity associated with larger Gram matrix multiplications in each local processing unit while supporting more total BS antennas. For example, for $C=32$ and $S=32$, we have  $B=1024$ BS antennas and achieve relatively high throughput. \revision{We also see that the CG detector achieves comparable or higher throughput than the ADMM detector for most cases due to its lower computational complexity.} \revision{Quite surprisingly, the ADMM beamformer can enable even higher performance than both ADMM and CG detectors.} In the $S=8$ case, for example, over $1$\,Gb/s of beamforming throughput can be achieved using a single ADMM iteration. This behavior is due to the fact that a single ADMM beamforming iteration (Algorithm~\ref{alg:admm_prec}) only requires local computations but no message passing, while ADMM and CG detectors require message passing. 
This indicates that, despite the optimizations described above, message passing latency still has a crucial effect on performance and further improvements in messaging may yield even higher data rates. 

\begin{rem}
We emphasize  these GPU cluster implementations serve as a proof-of-concept to showcase the efficacy and design scalability of DBP to large BS antenna arrays.
The achieved throughputs are by no means high enough for 5G wireless systems, which is mainly a result of the relatively high interconnect latency. 
{Nevertheless, we expect that DBP achieves throughputs in the  Gb/s regime if implemented on FPGA or ASIC clusters, which offer higher computing efficiency and lower interconnect latency (e.g., using Xilinx's GTY~\cite{gty} or Aurora protocols~\cite{aurora}) than that of GPU clusters.}
\end{rem}

\begin{rem}
{Power efficiency is another key aspect of practical BS designs. The thermal design power (TDP) of the Tesla K40 GPU used in our implementation is $235$\,W, leading to a maximum power dissipation of $C\times 235$\,W with $C$ fully-utilized GPUs. While this is a pessimistic power estimate, we expect that dedicated implementations on FPGA or ASIC will yield orders-of-magnitude better performance per watt.}
\end{rem}


\section{Conclusions}
\label{sec:conclusions}

\revision{We have proposed a novel decentralized baseband processing (DBP) architecture for massive MU-MIMO BS designs that mitigates interconnect and chip I/O bandwidth as well as complexity and signal processing bottlenecks.
DBP partitions the BS antenna array into independent clusters which perform channel estimation, data detection, and beamforming in a decentralized and parallel manner by exchanging only a small amount of consensus information among the computing fabrics.}
{The proposed data detection and beamforming algorithms achieve near-optimal error-rate performance at low complexity.  Furthermore, our simple consensus algorithms have low bandwidth requirements.}
Our GPU cluster implementation shows that the proposed method scales well to BS designs with thousands of antenna elements, and demonstrates that DBP enables the deployment of modular and scalable BS architectures for realistic massive MU-MIMO systems.

 \revision{We see numerous avenues for future work. A rigorous error-rate performance analysis of the proposed algorithms is an open research topic. The integration of our intra-cell decentralization schemes with inter-cell CoMP and C-RAN frameworks is a  direction worth to pursue in the future.} The development of decentralized algorithms for other 5G waveform candidates, such as SC-FDMA, FMBC, or GFDM, is left for future work. To alleviate the latency bottleneck, decentralized \emph{feedforward} architectures as in~\cite{JLCS17}  should be investigated for the downlink. \revision{Finally, an implementation of DBP on clusters with computing fabrics that have low interconnect latency and power consumption, such as FPGA or ASIC clusters, or heterogeneous or hybrid processors and accelerators for optimized workload deployment is part of ongoing work.}


\appendices
\section{Proofs}
\label{app:proofs}

\subsection{Proof of \fref{lem:D2simplification}}
\label{app:D2simplification}

We start by reformulating Step~E2 as follows:
\begin{align} \label{eq:D2step1}
 \bms^{(t+1)}  = \argmin_{\bms\in\complexset^U}  \, g(\bms) + \textstyle \sum_{c=1}^C {\frac{\rho}{2}} \vecnorm{\bms-\bmw^{(t)}_c}_2^2,
\end{align}
where we use the shorthand $\bmw^{(t)}_c=\bmz_c^{(t+1)}+\bmlambda_c^{(t)}$. Let $\bmv^{(t)}=\frac{1}{C}\bmw^{(t)}=\frac{1}{C}\sum_{c=1}^C\bmw^{(t)}_c$. Then, we can complete the square in the sum of the objective function of \eqref{eq:D2step1}, which yields
\begin{align*} 
\textstyle \sum_{c=1}^C \vecnorm{\bms-\bmw^{(t)}_c}_2^2 
 = &\, C \vecnorm{\bms}_2^2- \bms^HC\bmv^{(t)}-(\bmv^{(t)})^HC\bms \notag\\
  & \textstyle +\sum_{c=1}^C\vecnorm{\bmw^{(t)}_c}_2^2 
=  C\vecnorm{\bms - \bmv^{(t)}}_2^2 + K,
\end{align*}
where we define the constant $K= \sum_{c=1}^C\vecnorm{\bmw^{(t)}_c}_2^2 - C\vecnorm{\bmv^{(t)}}_2^2$. Since $K$ is independent of the  minimization variable $\bms$ in \eqref{eq:D2step1}, we obtain the equivalent minimization problem in~\eqref{eq:D2averaging}.

\subsection{Proof of \fref{lem:P2simplification}}
\label{app:P2simplification}

We start by reformulating Step (P2) as follows:
\revision{\begin{align}
 \bmz^{(t+1)} 
 & =  \argmin_{\bmz\in\complexset^{UC}, \vecnorm{\bms-\bD\bmz}_2\leq\varepsilon}\textstyle  \frac{1}{2} \vecnorm{\bmw_\text{all}^{(t)}-\bmz}_2^2,\label{eq:compactform}
\end{align}}where we define $\bD=\bOne_{1\times C}\kron\bI_U$, $\bmz^T=[\bmz_1^T \cdots\, \bmz_C^T]$, and $(\bmw_\text{all}^{(t)})^T=[(\bmw^{(t)}_1)^T \cdots\, (\bmw^{(t)}_C)^T]$ with $\bmw^{(t)}_c=\bH_c\bmx_c^{(t+1)}-\bmlambda_c^{(t)}$. 
Now, observe that the minimization problem~\eqref{eq:compactform} is the orthogonal projection of $\bmw_\text{all}^{(t)}$ onto the constraint $\vecnorm{\bms-\bD\bmz}_2\leq\varepsilon$.  We have the following closed-form expression for $\bmz^{(t+1)}$~\cite{studer2015demo}:
\begin{align*} 
  & \textstyle \bmw^{(t)}_\text{all} 
 +\! \max\!\left\{0,1\!-\!\frac{\varepsilon}{\|\bms-\bD\bmw^{(t)}_\text{all}\|_2}\right\} \!\bD^H(\bD\bD^H)^{-1}(\bms\!-\!\bD\bmw^{(t)}_\text{all}).
\end{align*}
We can simplify this expression using the identity
\begin{align*}
(\bD\bD^H)^{-1} &= ((\bOne_{1\times C}\kron\bI_U)(\bOne_{1\times C}\kron\bI_U)^H)^{-1}  \\
&= (C\kron\bI_U)^{-1}  = C^{-1}\bI_U
\end{align*}
{and $\bD^H\bD = \bOne_{C\times C}\kron\bI_U$. With these results, we obtain the following equivalent expression for $\bmz^{(t+1)}$
\begin{align*}
\bmw^{(t)}_\text{all} 
&  + \textstyle \max\!\left\{0,1-\frac{\varepsilon}{\|\bms-\bD\bmw^{(t)}_\text{all}\|}_2\right\} \times \\
& \qquad \qquad  \frac{1}{C} \left(\bD^H\bms- (\bOne_{C\times C}\kron\bI_U)\bmw_\text{all}^{(t)}\right)\!,
\end{align*}
which can be written using the per-cluster variables as}
\begin{align*}
 \bmz^{(t+1)}_c & = \textstyle \bmw^{(t)}_c + \max\!\left\{0,1-\frac{\varepsilon}{\|\bms-\vecv^{(t)}\|_2}\right\}\! \left( \frac{1}{C}\bms-\vecv^{(t)}\right)
\end{align*}
with $\vecv^{(t)}=\frac{1}{C}\bmw^{(t)}=\frac{1}{C}\sum_{c=1}^C\bmw^{(t)}_c;\bmw^{(t)}_c=\bH_c\bmx_c^{(t+1)}-\bmlambda_c^{(t)}$.


\section*{Acknowledgments}

The work of K.~Li, Y.~Chen, and J.~R.~Cavallaro was supported in part by the US National Science Foundation (NSF) under grants CNS-1265332, ECCS-1232274, ECCS-1408370, and CNS-1717218. The work of R.~Sharan and C.~Studer was supported in part by the US NSF under grants ECCS-1408006, CCF-1535897,  CAREER CCF-1652065, and CNS-1717559, and by Xilinx, Inc. The work of T.~Goldstein was supported in part by the US NSF under grant CCF-1535902 and by the US Office of Naval Research under grant N00014-17-1-2078.


\balance


\end{document}